\documentclass[twocolumn,superscriptaddress,showpacs,prl,amsmath,amstex,amssymb,citeautoscript,longbibliography]{revtex4-1}
\pdfoutput=1
\usepackage{natbib}
\usepackage[english]{babel}
\usepackage{letltxmacro}
\usepackage{latexsym}
\LetLtxMacro{\ORIGselectlanguage}{\selectlanguage}
\makeatletter
\DeclareRobustCommand{\selectlanguage}[1]{%
  \@ifundefined{alias@\string#1}
    {\ORIGselectlanguage{#1}}
    {\begingroup\edef\x{\endgroup
       \noexpand\ORIGselectlanguage{\@nameuse{alias@#1}}}\x}%
}
\newcommand{\definelanguagealias}[2]{%
  \@namedef{alias@#1}{#2}%
}
\makeatother
\definelanguagealias{en}{english}
\definelanguagealias{English}{english}
\usepackage{graphicx}
\usepackage{amsmath}
\usepackage{amsfonts}
\usepackage{amssymb}
\usepackage{bm}
\usepackage{color}
\usepackage[percent]{overpic}
\usepackage{soul} 
\usepackage{amssymb}
\usepackage{wasysym}
\usepackage{dsfont}
\usepackage{float}

\usepackage{hyperref}
\hypersetup{
    bookmarks=false,         
    unicode=false,          
    pdftoolbar=false,        
    pdfmenubar=true,        
    pdffitwindow=false,     
    pdfstartview={FitH},    
    pdftitle={Periodic orbits and quantum many-body scarring in constrained models: matrix product state approach},    
    pdfauthor={W. W. Ho},     
    pdfsubject={},   
    pdfcreator={},   
    pdfproducer={}, 
    pdfkeywords={many-body localization} {matrix elements} {disordered systems}, 
    pdfnewwindow=true,      
    colorlinks=true,       
    linkcolor=black,          
    citecolor=blue,        
    filecolor=magenta,      
    urlcolor=blue           
}

\newcommand{\be}{\begin{equation}}
\newcommand{\ee}{\end{equation}}
\newcommand{\bea}{\begin{eqnarray}}
\newcommand{\eea}{\end{eqnarray}}

\newcommand{\no}{\nonumber}
\newcommand{\mc}[1]{\mathcal{#1}}
\renewcommand{\vec}[1]{\boldsymbol{\mathbf{#1}}}

\usepackage{graphicx}
\usepackage[colorinlistoftodos]{todonotes}
\usepackage{verbatim}
\usepackage[normalem]{ulem}

\newcommand{\ket}[1]{\mbox{$ | #1 \rangle $}}
\newcommand{\bra}[1]{\mbox{$ \langle #1 | $}}

\begin{document}

\title{
Periodic orbits, entanglement and quantum many-body scars in constrained models: matrix product state approach 
}
\author{Wen Wei Ho}
\affiliation{Department of Physics, Harvard University, Cambridge, MA 02138, USA }
\author{Soonwon Choi}
\affiliation{Department of Physics, Harvard University, Cambridge, MA 02138, USA }
\author{Hannes Pichler}
\affiliation{ITAMP, Harvard-Smithsonian Center for Astrophysics, Cambridge, MA 02138, USA }
\affiliation{Department of Physics, Harvard University, Cambridge, MA 02138, USA }
\author{Mikhail D. Lukin}
\affiliation{Department of Physics, Harvard University, Cambridge, MA 02138, USA }

\date{\today}
\begin{abstract}
{ 
We analyze quantum dynamics of strongly interacting, kinetically constrained many-body systems. Motivated by recent experiments demonstrating surprising long-lived, periodic revivals after quantum quenches  in Rydberg atom arrays, we introduce a manifold of locally entangled spin states, representable by low-bond dimension matrix product states, and derive equations of motions for them using the time-dependent variational principle.  
We find that they feature isolated, unstable periodic orbits, which capture the recurrences and represent nonergodic dynamical trajectories. 
Our results provide a theoretical framework for understanding  quantum dynamics in a class of constrained spin models, which allow us to examine  the recently suggested  explanation of `quantum many-body scarring' \href{https://www.nature.com/articles/s41567-018-0137-5}{[Nature Physics {\bf 14}, 745-749 (2018)]}, and establish a {possible} connection to the corresponding   phenomenon in chaotic single-particle   systems.
}

\end{abstract}



\maketitle
 
{\it Introduction.} ---  
Understanding non-equilibrium dynamics in closed quantum many-body systems is of fundamental importance.
In ergodic systems, the eigenstate thermalization hypothesis (ETH) provides a means to describe their late-time, steady-state behavior by equilibrium statistical mechanics \cite{DeutschETH, SrednickiETH2, SrednickiETH3, Rigol2008, TarunETH}.
%
%
The few known exceptions to this paradigm include exactly solvable, integrable systems \cite{PhysRevLett.93.142002, PhysRevLett.98.050405, PhysRevLett.103.100403}, and strongly disordered, many-body localized systems, which feature extensive number of conservation laws \cite{Huse13, Serbyn13-1, RahulReview, AbaninMBLReview}. 
At the same time, the dynamics of equilibriation and thermalization 
 is  not as well understood. 
Concepts such as the ETH, while providing requirements for a system to eventually relax, do not unambiguously prescribe  the mechanism nor  the timescales on which this occurs;
interesting transient dynamics like prethermalization 
can occur 
\cite{PhysRevLett.93.142002, PhysRevLett.98.050405, PhysRevLett.103.100403, PhysRevLett.100.100601, PhysRevLett.98.180601, Gring1318, PhysRevLett.111.197203,PhysRevLett.116.120401, 2016AnPhy.367...96K,  PhysRevB.95.014112, Abanin2017, PhysRevLett.120.200601}.
%
%
%
%
Such non-equilibrium phenomena are generally challenging to analytically analyze \& simulate, and much progress has thus been spurred by  quantum simulation experiments in well-isolated, controllable many-body systems~\cite{Kinoshita2006, Kaufman794, Schreiber842, Rauer:2018,Meinert:2017dk,Labuhn:2016dp,Bernien2017,  Neyenhuise1700672, Zhang2017, Rainer2016, CTC,Choi16DTC, 2018arXiv180610169C}.

%
Recently, experiments on Rydberg atom arrays demonstrated surprising long-lived, periodic revivals after quantum quenches  \cite{Bernien2017}, with strong dependence of equilibriation timescales on the initial state.
Specifically, quenching from some unentangled product states, quick relaxation and thermal equilibriation of local observables was observed, typical of a chaotic, ergodic many-body system.
Conversely, quenching from certain other product states, coherent revivals  with a well-defined period were instead observed, which were  not seen to decay on the experimentally accessible timescales, a distinctively nonergodic  dynamical behavior. 
Most surprisingly, these strikingly different behavior resulted from initial states that are
all highly excited with similar, extensive energy densities, and are hence
 indistinguishable from a thermodynamic standpoint.  
The apparent simplicity of the special, slowly thermalizing initial states' dynamics -- {\it periodic, coherent many-body oscillations} -- therefore brings to question whether they can be understood  in a simple, effective picture. 
In fact, recent theoretical work~\cite{QMBS} suggested an intriguing analogy of the  oscillations with the phenomenon of quantum scarring in chaotic single-particle  systems, where 
a quantum particle shows  {similarly long-lived} periodic revivals when launched along weakly unstable, periodic orbits of the underlying classical model \cite{Heller}. 
%
However, to date, a firm  connection to the theory of single-particle quantum scars~\cite{Heller} has not been established.

\begin{figure}[t] \includegraphics[width=0.5\textwidth]{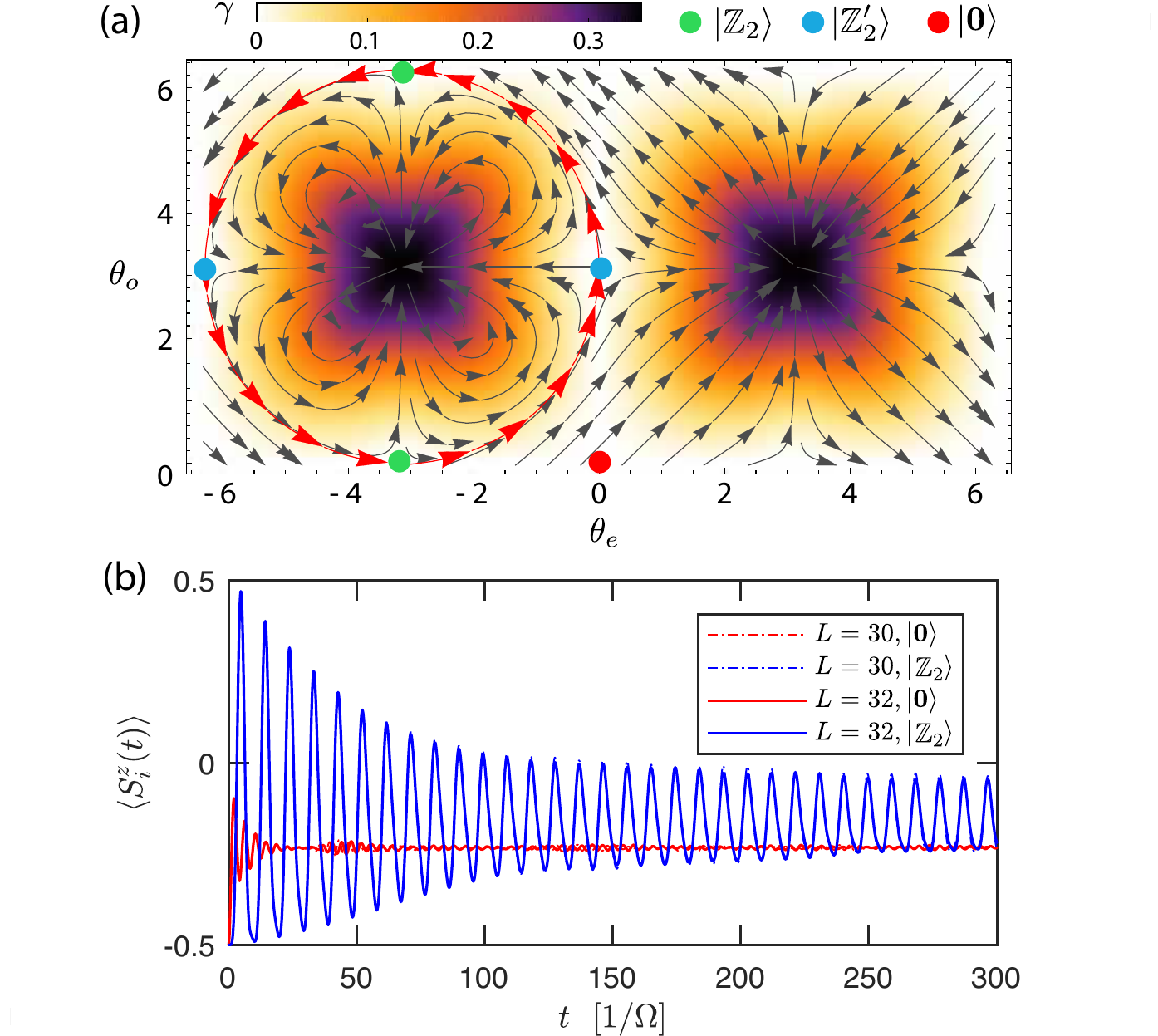}
\caption{(a) Flow diagrams  of $\dot{\theta}_e(t), \dot{\theta}_o(t)$  for the model (\ref{eqn:H}) with $s$\,$=$\,$1/2$. The color map gives the error $\gamma$, (\ref{eqn:gamma}). There is an isolated, unstable periodic orbit (red curve) describing oscillatory motion between $|\mathbb{Z}_2\rangle$ (green dot) and  $|\mathbb{Z}'_2\rangle$ (blue dot), with numerically extracted  period $T$\,$\approx$\,$2 \pi$\,$\times$\,$1.51$\,$\Omega^{-1}$. Conversely,   motion from $|\mathbf{0}\rangle$ (red dot) proceeds towards a saddle point where the error is large.
%
(b): Dynamics of local observable $S^z_i(t)$.
%
 %
There are persistent, coherent oscillations in the local observable for  $|\mathbb{Z}_2\rangle$ with similar period, while $|\mathbf{0}\rangle$  instead shows quick relaxation and equilibriation towards a thermal value predicted by ETH \cite{appendix}. 
}
\label{fig:FlowDiagram1}
\end{figure}


In this Letter, we develop a theoretical framework to  analyze the quantum dynamics of a family of constrained spin models, which display similar phenomenology of long-lived periodic revivals from  certain special initial states. 
%
Specifically, 
we introduce a manifold of simple, locally entangled states respecting the constraints, representable by a class of low bond dimension matrix product states (MPS), and derive equations of motions (EOM) for them using the time-dependent variational principle (TDVP) \cite{TDVP1, TDVP2}.
We find that these EOM support isolated, unstable, periodic orbits.  
By quantifying the accuracy of this effective description, we show  that these closed orbits indeed 
capture the persistent recurrences, 
and hence signal slow relaxation of local observables, a form of weak ergodicity breaking in dynamics, see Fig.~\ref{fig:FlowDiagram1}(a,b). 
%
Furthermore, since the TDVP generates a Hamiltonian flow in the phase space parametrizing this (weakly entangled) manifold, one can associate 
our approach with a
 generalized ``semiclassical'' description of many-body dynamics in constrained Hilbert spaces.
Our 
finding of periodic orbits in this description is therefore suggestive in {establishing} the connection to the theory of quantum scarring of single-particle systems of Heller~\cite{Heller}. 

{\it Kinetically constrained spin models.} ---  
We consider a family of interacting, constrained spin models and demonstrate that they show atypical thermalization behavior for certain initial states.
 Consider a  chain of $L$ spin-$s$ particles on a ring,
with   Hamiltonian 
\begin{align}
H =   \Omega  \sum_{i}   \mathcal{P} S^x_i \mathcal{P}.
\label{eqn:H}
\end{align}
Here, a basis on each site $i$ is spanned by eigenstates 
$|n\rangle_i$ of $S^z_i$\,$+$\,$s$\,$\mathbb{I}_i$, with $n$\,$=$\,$0,\cdots$\,$,2s$, and 
$S_i^x$ is the spin-$s$ operator in the $x$-direction.
%
 The projector $\mathcal{P}$\,$=$\,$\prod_i$\,$\mathcal{P}_{i,i+1}$  is a product of commuting local projectors $\mathcal{P}_{i,i+1}$\,$=$\,$\mathbb{I}_{i}$\,$\otimes$\,$\mathbb{I}_{i+1}$\,$-$\,${Q}_i$\,$\otimes$\,${Q}_{i+1}$, with ${Q}_i$\,$=$\,$\mathbb{I}_{i}$\,$-$\,$P_i$ and $P_i$\,$=$\,$\ket{0}_i\bra{0}_i$, and constrains   dynamics to a subspace where at least one of  two  neighboring  spins  is in the state $\ket{0}$, which has dimensionality  $d$\,$\sim$\,$((1$\,$+$\,$\sqrt{8s+1})/2)^L$. 
%
%
When $s$\,$=$\,$1/2$, Eqn.~(\ref{eqn:H}) 
effectively models the experimental setup of \cite{Bernien2017}, 
where the 
constraint stems from the Rybderg blockade mechanism (see also~\cite{RS, Lesanovsky1,Lesanovsky0, Lesanovsky2, Lesanovsky3}).
%
%


\begin{figure}[t]
\includegraphics[width=0.5\textwidth]{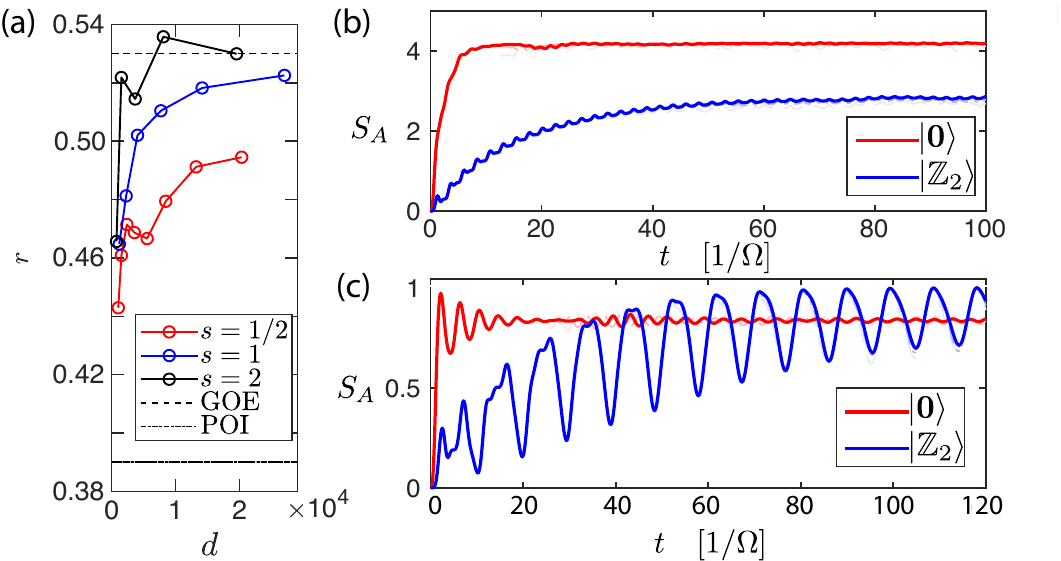}
\caption{(a) Level spacing statistics in the momentum-zero, inversion-symmetric sector. Plotted is the $r$-statistics defined by the average of $r_n$\,$=$\,$\frac{\text{min}(s_n,s_{n-1})}{\text{max}(s_n,s_{n-1}) }$ where $s_n$\,$=$\,$E_{n+1}$\,$-$\,$E_n$. There is a clear albeit slow trend with Hilbert space dimension $d$ towards Wigner-Dyson statistics in the GOE class, indicated by $r$\,$\approx$\,$0.53$, away from the integrable Poissonian (POI) limit of $r$\,$\approx$\,$0.39$ (for discussion of the slow convergence, see \cite{VA,  2018arXiv180610933T}). (b,c) Growth of entanglement entropy  $S_A$ following quenches from the $|\bm0\rangle $ and $|\mathbb{Z}_2\rangle$ states, of subregions $A$ being (b) six contiguous sites, (c) a single-site, for the $s$\,$=$\,$1/2$ model. Total system size is $L$\,$=$\,$30$.
%
%
}
\label{fig:system}
\end{figure}


The Hamiltonian (\ref{eqn:H}) has a simple interpretation: each spin  rotates freely about the $x$-axis if both its neighbors  are in the state $\ket{0}$, while its dynamics is frozen otherwise.
Despite its apparent simplicity, the Hamiltonian is  nonintegrable and quantum chaotic, as seen in Fig.~\ref{fig:system}(a) from level repulsion in the energy eigenspectrum.  
The chaotic nature of the system is expected to govern the nonequilibrium dynamics arising from a quantum quench. For example, consider ``simple'', unentangled initial states, specifically product states in the $z$-basis  that satisfy the constraints. All these states have the property that they have the same energy density under  (\ref{eqn:H}), corresponding to that of the infinite-temperature thermal state, and are hence thermodynamically indistinguishable. 
Under time evolution, one would expect a quick relaxation of local observables (on the timescale $t_r$\,$\sim$\,$\Omega^{-1}$) to infinite-temperature ensemble values \cite{appendix}, in accordance with ETH predictions~\cite{DeutschETH, SrednickiETH1, SrednickiETH2, SrednickiETH3, Quench1, Quench2, Quench3}. %
This behavior is indeed observed generically, as demonstrated previously~\cite{Lesanovsky1, Lesanovsky0, Lesanovsky2, Lesanovsky3}, and also in Fig.~\ref{fig:FlowDiagram1}(b) for
the local observable $ S^z_i(t)$ from the initial state $\ket{\mathbf 0}$\,$=$\,$\otimes_{i=1}^L\ket{0}_i$ ($s$\,$=$\,$1/2$).
 However,   time evolution of the initial state $|\mathbb{Z}_2\rangle$\,$\equiv$\,$\otimes_{i=1}^{L/2}$\,$\ket{0}_{2i-1}$\,$\ket{2s}_{2i}$ does not follow this expectation. As shown in  Fig.~\ref{fig:FlowDiagram1}(b), the same observable instead unexpectedly exhibits long-lived, coherent oscillations with a well-defined period $T$\,$\approx$\,$2\pi$\,$\times$\,$1.51$\,$\Omega^{-1}$. 
Furthermore, it does not relax to, nor oscillate about, the thermal value expected from ETH, at least on numerically accessible timescales and system sizes. 

This striking departure from   generic behavior is also reflected in the growth of  entanglement entropy (EE)   (Fig.~\ref{fig:system}(b,c)). While for generic initial states EE essentially grows linearly and quickly saturates to that of a random state \cite{appendix}, this is not the case for $|\mathbb{Z}_2\rangle$.
In particular, the single-site EE drops periodically, indicating that each spin is repeatedly partially disentangling itself from the rest of the chain.
%
This 
 tantalizingly  hints that the motion for the $|\mathbb{Z}_2\rangle$ state  lies 
within a low-entanglement manifold of the Hilbert space, thereby possibly allowing for a simple, effective description of dynamics.
%


\begin{figure}[t]
\includegraphics[width=0.5\textwidth]{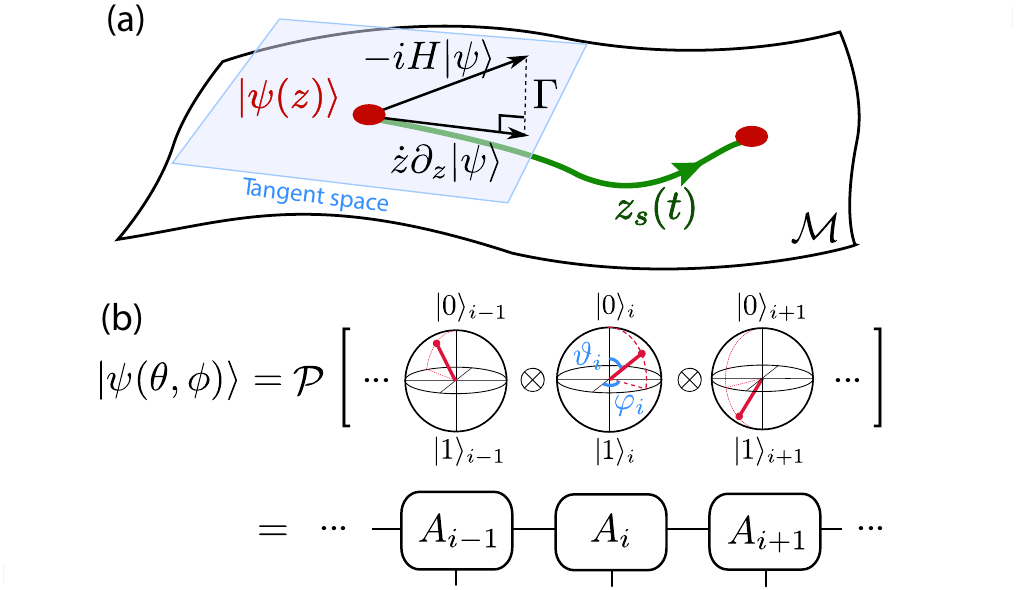}
\caption{
(a) Geometrical depiction of the TDVP over a manifold of states $|\psi(\bm z)\rangle$ parameterized by  $\bm z$. The instantaneous motion $-i H |\psi (\bm z) \rangle$ is projected onto the tangent space at the point, leading to motion on the manifold (green trajectory). 
The norm of the vector orthogonal to the manifold, $\Gamma$\,$=$\,$\gamma$\,$\sqrt{L}$ (c.f.~Eqn.~(\ref{eqn:gamma})), is a measure of its accuracy. 
(b) MPS representation of states $|\psi(\bm{\theta},\bm{\phi})\rangle$ (c.f.~Eqn.~(\ref{eqn:MPS})) used. 
}
\label{fig:PathIntegral}
\end{figure}
 
{\it  Equations of motion from the TDVP.} ---
Motivated by these considerations,  we  analyze the dynamics of the system using the TDVP  on a suitable variational manifold of simple, low entanglement states. 
 For concreteness, we focus first on $s$\,$=$\,$1/2$.
Starting from classical spin configurations, i.e.~products of unentangled coherent states $\otimes_i |\vartheta_i, \varphi_i\rangle$\,$:=$\,$\otimes_i [ \cos(\vartheta_i/2)\ket{0}_i-ie^{i\varphi_i}\sin(\vartheta_i/2)|1\rangle_i ]$, we construct states that respect the constraints set by $\mathcal{P}$, by explicitly projecting out neighboring excitations,
\begin{align}
|\psi(\bm{\vartheta},\bm{\varphi}) \rangle = \mathcal{P} \bigotimes_{i} |(\vartheta_i, \varphi_i)\rangle,
\label{eqn:psi}
\end{align}
which is akin to a Gutzwiller projection to the constrained subspace \cite{Gutzwiller, appendix}, see Fig.~\ref{fig:PathIntegral}(b). Importantly, (\ref{eqn:psi}) is weakly entangled, and can be written as a particular matrix product state (MPS) with bond dimension $D$\,$=$\,$2$ \cite{VidalMPS, appendix}. 
%
%
%
We   find it convenient to normalize \eqref{eqn:psi} and change to new variables   $(\bm{\vartheta}, \bm{\varphi})$\,$\to$\,$(\bm{ \theta}, \bm{\phi})$ via a non-linear mapping  \cite{appendix}, such that $\ket{\psi(\bm{ \vartheta}, \bm{ \varphi} )}/||\psi(\bm{ \vartheta}, \bm{ \varphi} )||$\,$=$\,$|\psi(\bm{ \theta}, \bm{ \phi} ) \rangle$, so that    the MPS repesentation  is given by
%
\begin{align} \label{eqn:MPS}
|\psi(\bm{ \theta}, \bm{ \phi} ) \rangle &= \text{Tr}(A_1 A_2 \cdots A_L),
\nonumber \\
A_i(\theta_i, \phi_i) & = \begin{pmatrix} 
P_i | (\theta_i, \phi_i) \rangle & Q_i | (\theta_i, \phi_i) \rangle \\
 |0\rangle_i & 0 \end{pmatrix},
\end{align}
and  $|(\theta_i, \phi_i)\rangle$\,$=$\,$e^{i \phi_i s}e^{i \phi_i S^z_i} e^{-i \theta_i S^x_i} |0\rangle_i$, which is  normalized in the thermodynamic limit $L$\,$\rightarrow$\,$\infty$ (see too \cite{PhysRevLett.106.025301, PhysRevLett.108.105301}). 
The generalization of (\ref{eqn:MPS}) to   spin-$s$   then simply consists of replacing the appropriate operators and states with the spin-$s$ analogs.
%

The TDVP respects conservation laws, and in particular  conserves the energy of the Hamiltonian (\ref{eqn:H}) \cite{TDVP1, TDVP2,Leviatan2018, appendix}. On this general ground, we obtain that $\bm{\dot{\phi}}$\,$=$\,$0$, and can set $\bm{\phi}$\,$=$\,$0$, which is obeyed for initial product states in the $z$-basis \cite{appendix}. 
Furthermore, to describe the motions of the $|\bm{0}\rangle$ and $|\mathbb{Z}_2\rangle$ states, it suffices to focus on the submanifold of states with a two-site translational symmetry, i.e.~$\theta_i$\,$=$\,$\theta_{i+2}$.
The TDVP-EOM are obtained by projecting the instantaneous motion of the quantum system  onto the tangent space of the  variational manifold (Fig.~\ref{fig:PathIntegral}(a)), and read $\sum_\mu \bm{\dot \theta}_\mu  \langle \partial_{\bm{\theta}_\nu}  \psi | \partial_{\bm{\theta}_\mu} \psi \rangle$\,$=$\,$-i  \langle  \partial_{\bm{\theta}_\nu} \psi |  H |\psi\rangle$, for $\mu$\,$\in$\,$\{o,e\}$ (standing for even(e) and odd(o) sites).
A lengthy but straightforward calculation \cite{appendix} yields closed-form, analytic expressions: 
 $\dot\theta_{\rm e}(t)$\,$=$\,$f(\theta_{\rm e}(t),\theta_{\rm o}(t))$ and $\dot\theta_{\rm o}(t)$\,$=$\,$f(\theta_{\rm o}(t),\theta_{\rm e}(t))$, with
\begin{align}\label{eqn:FlowEq}
f(x,y) &= \Omega  \left[ 1 - \cos^{4s-2} \left(\frac{x}{2} \right) + \cos^{4s-2}\left(\frac{x}{2} \right)\cos^{2s}\left(\frac{y}{2} \right) \right. \nonumber \\
& \left. + 2s \sin\left(\frac{x}{2}\right) \cos^{6s-1}\left( \frac{x}{2} \right)\tan \left(\frac{y}{2} \right)         \right].
\end{align}

\begin{figure}[t]
\includegraphics[width=0.5\textwidth]{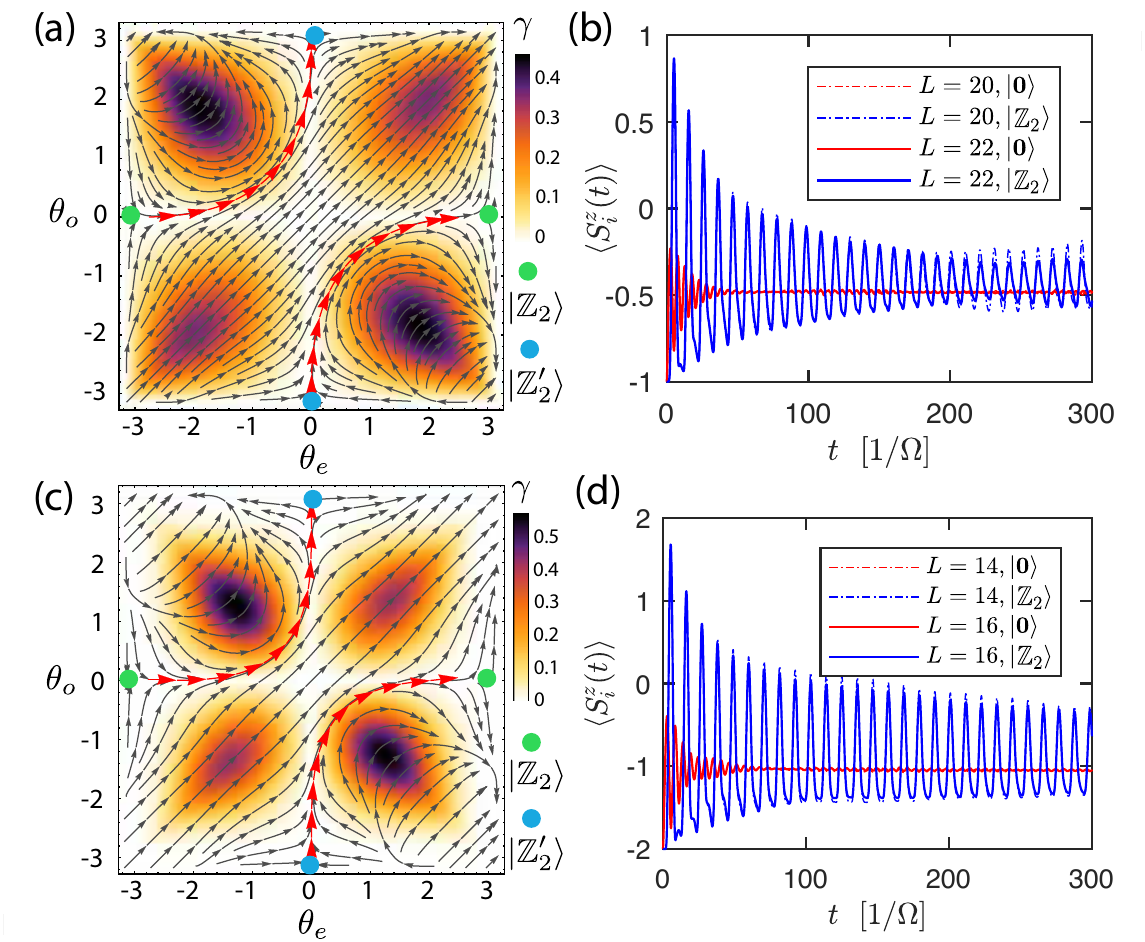}
\caption{(a,c) Flow diagrams  (\ref{eqn:FlowEq}) and error $\gamma$ for  (a) $s$\,$=$\,$1$, (c) $s$\,$=$\,$2$.
The indicated periodic orbits (red curves) have periods (a) $T$\,$\approx$\,$2 \pi$\,$\times$\,$1.64$\,$\Omega^{-1}$, and (c) $T$\,$\approx$\,$2\pi$\,$\times$\,$1.73$\,$\Omega^{-1}$. Note that points $\theta_{o/e}$\,$=$\,$\theta_{o/e}$\,$\pm$\,$2\pi$ are identified.
%
(b,d) Relaxation of local observable $ S^z_i(t) $ for (b) $s$\,$=$\,$1$, (d) $s$\,$=$\,$2$. One sees, similarly to Fig.~\ref{fig:FlowDiagram1}, quick relaxation of the $|\bm{0}\rangle$ state toward a thermal value predicted by ETH \cite{appendix}, while persistent  oscillations for $|\mathbb{Z}_2\rangle$, with similar periods in (a,c).
}
\label{fig:orbit}
\end{figure}

These EOM are coupled, nonlinear equations. Yet, remarkably, we find that for each spin-$s$, there is an isolated, unstable, periodic orbit $\mathcal{C}$, as seen in the corresponding flow diagrams for $s$\,$=$\,$1/2$ in Fig.~\ref{fig:FlowDiagram1}(a), and $s$\,$=$\,$1,2$, in Fig.~\ref{fig:orbit}(a,c).
%
%
 Furthermore, $\mathcal{C}$ includes the points $(\theta_e,\theta_o)$\,$=$\,$(\pi,0)$, and $(0,-\pi)$ (modulo $2\pi$), corresponding to $\ket{{\mathbb{Z}_2}}$ and its counterpart $\ket{{\mathbb{Z}'_2}}$\,$=$\,$\otimes_{i=1}^{L/2}\ket{0}_{2i}\ket{2s}_{2i-1}$ respectively. Thus, the  EOM describe continual oscillations between these two product states (akin to a quantum Newton's cradle! [see also \cite{Kinoshita2006}]), which is manifestly an athermal, nonergodic behavior \footnote{Note that due to the parametrization, $(\theta_e$\,$,$\,$\theta_o)$\,$=$\,$(\pi/2,0)$ gives the same state as $(\theta_e,\theta_o)$\,$=$\,$(\pi/2,-\pi)$. There is additionally a coordinate singularity at each point.
}.
 The periods of oscillations from the EOM can be determined by numerical integration of Eq.~\eqref{eqn:FlowEq}, and the extracted values match excellently with
those from  numerical simulations of local observables such as  $S^z_i(t)$, see Fig.~\ref{fig:FlowDiagram1}(b)  and Fig.~\ref{fig:orbit}(b,d).   
This  already indicates that the variational manifold (\ref{eqn:MPS}) is well suited to capture central aspects of the exact quantum dynamics.

To further corroborate this fact, we  quantify the error in TDVP evolution as the instantaneous rate at which the   state evolving under the full  Hamiltonian leaves the variational manifold (see Fig.~\ref{fig:PathIntegral}, \cite{TDVP1, TDVP2}), given by 
\begin{align}
\gamma(\vec{\theta} ) =  || (i H + \dot{\vec{\theta}}\partial_{\vec{\theta}}) 
  |\psi(\vec{\theta} )\rangle 
 ||/\sqrt{L},
\label{eqn:gamma}
\end{align} 
where we have normalized it to be an intensive quantity. 
The numerically integrated error rates  around the closed orbits $\epsilon_{\mathcal C}$\,$=$\,$\oint_{\mathcal C }$\,$\gamma(\theta_e(t)$\,$,$\,$\theta_o(t))dt$  yield $\epsilon_{ \mathcal C }$\,$\approx$\,$0.17, 0.32, 0.41$
 for $ s$\,$=$\,$1/2,$\,$1,$\,$2$ respectively, which are small values   compared to neighboring trajectories \cite{appendix}, illustrating that  $\mathcal{C}$ is indeed a good approximation to exact quantum dynamics.
%
We stress that the ability to capture the key features of some dynamics of a chaotic many-body system within a low entanglement manifold is remarkable.
This is in contrast to generic expectations; for example, the trajectory beginning at $(\theta_e,\theta_o)$\,$=$\,$(0,0)$ for $s$\,$=$\,$1/2$, (i.e.~the $|\mathbf{0}\rangle$ state), instead traces out a path that terminates in a saddle point  where  $\gamma$ is large (see Fig.~\ref{fig:FlowDiagram1}(a)), indicating that this low entanglement manifold is   unable to capture the large growth of entanglement from this state, as expected in a thermalizing system.
%

{\it Discussion.} ---   Our effective description of the persistent oscillations seen in the many-body systems (\ref{eqn:H}), in terms of isolated, unstable orbits, provides a framework to analyze a possible connection with the phenomenon of quantum scarring in single-particle chaotic systems \cite{Heller}.
There, special, weakly unstable classical orbits of a single-particle, characterized by the condition $\lambda T$\,$<$\,$1$ (where $T$ is the period of the orbit and $\lambda$ the average Lyapunov exponent about the orbit)  play a central role:  
the persistent revivals and slow decay of a Gaussian wavepacket (a quatum particle) launched along such an orbit give rise to a statistically significant enhancement of certain wavefunctions' probability densities about these orbits, above that expected of Berry's conjecture \cite{BerryConjecture}.
Indeed, the apparent similarity between these phenomena, and atypical signatures in the ergodic properties of certain many-body eigenstates of the $s$\,$=$\,$1/2$ model (\ref{eqn:H}) tied to the long-lived oscillations, motivated the recently proposed explanation in terms of quantum many-body scars  \cite{QMBS, 2018arXiv180610933T}.
Our work provides a way to make such an analogy firmer: 
even though our variational manifold encompasses states that explicitly include quantum entanglement, the TDVP-EOM describe a Hamiltonian flow in the corresponding phase space \cite{TDVP1,TDVP2,PathIntegralEntangled,Hallam}, and thus offer a notion of a ``semiclassical trajectory'' through the many-body Hilbert space.  
A natural extension of the  condition $\lambda T$\,$<$\,$1$ characterizing the instability of orbits 
is then the leakage out of the manifold $\epsilon_{ \mathcal C }$\,$=$\,$\oint_{ \mathcal C }\gamma(\vec{\theta}) dt$\,$<$\,$1$; it would be interesting to relate this quantity to the Lyapunov exponent of the EOM \cite{Hallam}.
Furthermore,  the effect of these orbits on the nature of many-body eigenstates deserve further study; however this has to be done while contending with the thermodynamic limit, a notion absent in the single-particle scenario.

Finally, we note that the equations of motion we obtained can also be understood as the leading order, saddle-point evaluation of a path integral for the constrained spin systems (\ref{eqn:H}).
 In particular, the manifold of states $|\psi (\bm{\theta}, \bm{\phi})\rangle$  is dense and supports a resolution of the identity on the constrained space, 
with an appropriate measure $\mu(\bm{ \theta}, \bm{\phi} )$ (see \cite{appendix}), allowing the construction of a Feynman path integral \cite{RevModPhys.20.367, Weinberg, AZee, Altland, PathIntegralEntangled}. 
The TDVP-EOM extremize the action functional with the Lagrangian $\mc{L}$\,$=$\,$i \langle \psi| \partial_{\bm{\theta}} \psi\rangle  \dot{\bm{\theta}}$\,$+$\,$i \langle \psi| \partial_{\bm{\phi}} \psi\rangle  \dot{\bm{\phi}}
$\,$-$\,$\langle \psi| {H} | \psi\rangle$, which evaluates (for $s$\,$=$\,$1/2$) to:
\begin{align}
&\mc{L}=\sum_{i}K_i(\vec{\theta}) [ \sin^2 \!\left(\!\frac{\theta_i}{2} \!\right)\!\dot\phi_i+ \frac{\Omega}{2} \cos \!\left(\!\frac{\theta_{i+1} }{2} \!\right) \sin \left(\theta_i \right) \cos (\phi_i) ], \no
\end{align}
where $K_i(\vec{\theta})$ is given in \cite{appendix}.
This formulation provides a framework, which can be used to systematically recover quantum dynamics from the saddle-point limit, by including higher-order corrections, i.e.~fluctuations.


{\it Conclusion.} --- 
In this Letter, we introduced and analyzed the dynamics of a family of     constrained spin models which show atypical thermalization behavior -- long-lived, coherent revivals from certain special initial states, similar to recent quench experiments in a quantum simulator of Rydberg atoms.
%
%
%
We derived an effective  description of these systems  
 in terms of equations of  motion for dynamics of locally entangled spins and found that they host isolated, unstable, periodic orbits, which correspond to   long-lived recurrences at the  quantum many-body level. Our results 
establish a {possible} connection to quantum scarring in single-particle chaotic systems, and suggest a framework for a  generalization of the theory of quantum scars {by Heller \cite{Heller}, which is intimately tied to unstable periodic orbits,} to the many-body case.

While our analysis demonstrates  that the phenomenology of stable, long-lived oscillations from special initial states extends to a number of interacting, constrained   models, one of the most important outstanding questions is related to their physical origin and the sufficient conditions for their existence. A complementary Letter \cite{VA} demonstrates that  these models possess important features resembling ergodic systems that are close to integrability, and that these features can be enhanced by non-trivial deformations of the  Hamiltonian. In \cite{appendix}, we  show that our variational description of the periodic dynamics is able to capture the effect of these deformations by making the corresponding error $\gamma$ smaller. 
 While it is currently unclear if this near-integrable-like behavior is directly related to,  required for, or follows from the existence of scar-like dynamics (see however recent work \cite{perfectScars} exploring the role
of deformations on stabilizing the periodic dynamics), 
these observations as well as the framework presented here provide both theoretical foundations and important physical insights on which future studies of quantum dynamics can be based upon.

 \begin{acknowledgements} {\it Acknowledgments.} ---  We thank  V.~Khemani, A.~Chandran,  D.~Abanin, A.~Vishwanath, D.~Jafferis, E.~Demler, J.~Nieva-Rodriguez, V.~Kasper and E.~Heller for useful discussions.   This work was supported through the National Science Foundation (NSF), the Center for Ultracold Atoms, the Air Force Office of Scientific Research via the MURI, and the Vannevar Bush Faculty Fellowship. H.P.~is supported by the NSF through a grant for the Institute for Theoretical Atomic, Molecular, and Optical Physics at Harvard University and the Smithsonian Astrophysical Observatory. W.W.H.~is supported by the Gordon and Betty Moore Foundation's EPiQS Initiative   Grant No. GBMF4306. \end{acknowledgements}

 \bibliography{refs}

\end{document}


\title{
{Supplemental Material:  \\ Periodic orbits, entanglement and quantum many-body scars in constrained models: matrix product state approach }
}
\author{Wen Wei Ho}
\affiliation{Department of Physics, Harvard University, Cambridge, MA 02138, USA }
\author{Soonwon Choi}
\affiliation{Department of Physics, Harvard University, Cambridge, MA 02138, USA }
\author{Hannes Pichler}
\affiliation{ITAMP, Harvard-Smithsonian Center for Astrophysics, Cambridge, MA 02138, USA }
\affiliation{Department of Physics, Harvard University, Cambridge, MA 02138, USA } 
\author{Mikhail D. Lukin}
\affiliation{Department of Physics, Harvard University, Cambridge, MA 02138, USA }

\date{\today}
\maketitle

In this supplemental material, we  (i) provide details on normalizing the `Gutzwiller projected' variational state via a non-local mapping, (ii) derive the effective equations of motion using the time-dependent variational principle (TDVP), as well as the error $\gamma$, (iii) derive the measure $\mu(\bm{\theta}, \bm{\phi})$ for the purposes of writing a resolution of the identity on the constrained space, and hence a path integral, (iv) explain what it means to thermalize in the constrained space, and (v) repeat the TDVP calculations for the deformed model of \footnote{V.~Khemani, C.~R.~Laumann, and A.~Chandran, ``Signatures of integrability in the dynamics of Rydberg-blockaded chains'', ArXiv e-prints (2018), arXiv:1807.02108 [cond-mat.str-el].}.

\section{I. Normalizing the `Gutzwiller projected' state}
In this section we show that for spin $s=1/2$, the `Gutwziller projected' state 
\begin{align}
 | \psi(\bm \vartheta, \bm \varphi)\rangle & = \mathcal{P} \bigotimes_i | (\vartheta_i,\varphi_i) \rangle,  
\end{align}
where $  |(\vartheta_i, \varphi_i)\rangle  =e^{i {\varphi_i}/{2} }e^{i \varphi_i S^z_i} e^{-i \vartheta_i S^x_i } |0\rangle_i =  \cos( \vartheta_i/2) |0\rangle_i - i e^{i \varphi_i} \sin(\vartheta_i/2) | 1\rangle_i$ (a spin-coherent state), can be normalized and written  explicitly as a bond dimension two matrix product state (MPS), i.e.~
\begin{align}
& | \psi(\bm \theta, \bm \phi)\rangle \equiv \frac{| \psi(\bm \vartheta, \bm \varphi)\rangle }{||| \psi(\bm \vartheta, \bm \varphi)\rangle ||}   = \text{Tr}(A_1 A_2 \cdots A_L), \nonumber \\
& A_i(\theta_i, \phi_i)  = \begin{pmatrix}
P_i |(\theta_i, \phi_i)\rangle & Q_i |(\theta_i, \phi_i)\rangle \\
 | 0 \rangle_i & 0
\end{pmatrix},
\end{align}
via a non-local mapping $(\bm \vartheta, \bm \varphi) \to (\bm \theta, \bm \phi)$. This is the form of the variational state used in the TDVP calculation.
In the above, 
\begin{align}
|(\theta_i, \phi_i)\rangle =e^{i {\phi_i}/{2} }e^{i \phi_i S^z_i} e^{-i \theta_i S^x_i } |0\rangle_i =  \cos( \theta_i/2) |0\rangle_i - i e^{i \phi_i} \sin(\theta_i/2) | 1\rangle_i
\end{align}
 is (another) spin-coherent state,  
$
\mathcal{P} = \prod_i \mathcal{P}_{i,i+1}
$
 the projector onto the constrained subspace, with $\mathcal{P}_{i,i+1}$  a local projector defined as
$
\mathcal{P}_{i,i+1} = \mathbb{I}_i\otimes \mathbb{I}_{i+1} - Q_i \otimes Q_{i+1},
$ and $Q_i = \mathbb{I}_i - P_i$, $P_i = |0\rangle_i \langle 0 |_i$.

\subsection{MPS representation }
We start by writing $| \psi(\bm \vartheta, \bm \varphi)\rangle = \mathcal{P} \bigotimes_i | (\vartheta_i,\varphi_i) \rangle$ as a bond dimension two MPS. 
This is possible because $\mathcal{P}$ can be cast as a matrix product operator bond dimension two. 
To derive this, we iterative apply the projector $\mathcal{P}_{i,i+1}$ on each pair of sites starting from one end of the chain. Letting $a_i = \cos(\vartheta_i/2), b_n = -i e^{i \varphi_i} \sin(\vartheta_i/2)$, we have
\begin{widetext}
\begin{align}
| \psi(\bm \vartheta, \bm \varphi)\rangle & = \mathcal{P} \bigotimes_i^L | (\vartheta_i,\varphi_i) \rangle \nonumber\\
& = \left( \prod_{j=1}^L \mathcal{P}_{j,j+1} \right)  \bigotimes_i^L | (\vartheta_i,\varphi_i) \rangle \nonumber \\
& = \left( \prod_{j=2}^L \mathcal{P}_{j,j+1} \right) (a_1 a_2 |0\rangle_1 |0\rangle_2 + a_1 b_2|0\rangle_1 |1\rangle_2 + b_1 a_2 |1\rangle_1 |0\rangle_2 ) \bigotimes_{i=3}^L | (\vartheta_i,\varphi_i) \rangle \nonumber \\
& = \left( \prod_{j=2}^L \mathcal{P}_{j,j+1} \right) 
 \text{Tr}\left[
\begin{pmatrix} a_1 |0\rangle_1 & b_1 |1\rangle_1 \\ a_1 |0\rangle_1 & 0 \end{pmatrix}
\begin{pmatrix} a_2 |0\rangle_2 & b_2 |1\rangle_2 \\ a_2 |0\rangle_2 & 0 \end{pmatrix}
\right] \bigotimes_{i=3}^L | (\vartheta_i,\varphi_i) \rangle \nonumber \\ 
& =  \left( \prod_{j=2}^L \mathcal{P}_{j,j+1} \right)  \text{Tr} \left[ \mc A(a_1, b_2) \mc A(a_2,b_2) \right] \bigotimes_{i=3}^L | (\vartheta_i,\varphi_i) \rangle,
\end{align}
\end{widetext}
and where 
\begin{align}
\mc A(a_n, b_n) = \begin{pmatrix} a_n |0 \rangle_n & b_n |1 \rangle_n \\ a_n |0\rangle_n & 0  \end{pmatrix}.
\end{align}
In the above, the more conventional representation would entail a decomposition into basis states, i.e.~
\begin{align}
\mc A(a_n, b_n) =  \sum_s \mc A^s(a_n, b_n) | s\rangle,
\end{align}
where $s = 0,1$ and 
\begin{align}
& \mc A^0(a_n, b_n) = \begin{pmatrix} a_n  & 0 \\ a_n  & 0  \end{pmatrix}, \qquad \mc A^1(a_n, b_n) = \begin{pmatrix} 0  & b_n  \\ 0  & 0  \end{pmatrix}.
\end{align}

For the induction step we assume that
\begin{align}
| \psi(\bm \vartheta, \bm \varphi)\rangle & = \left( \prod_{j=k}^L \mathcal{P}_{j,j+1} \right)  \text{Tr} \left[ \prod_{n=1}^k A(a_n, b_n) \right] \bigotimes_{i=k+1}^L | (\vartheta_i,\varphi_i) \rangle
\end{align}
and it is then easy to show
\begin{align}
| \psi(\bm \vartheta, \bm \varphi)\rangle & = \left( \prod_{j=k+1}^L \mathcal{P}_{j,j+1} \right)  \text{Tr} \left[ \prod_{n=1}^{k+1} \mc A(a_n, b_n) \right] \bigotimes_{i=k+2}^L | (\vartheta_i,\varphi_i) \rangle.
\end{align}
Therefore, we have that
\begin{align}
\boxed{  | \psi(\bm \vartheta, \bm \varphi)\rangle  =  \text{Tr} [ \mc A(a_1, b_1) \mc A(a_2, b_2) \cdots \mc A(a_L, b_L) ]. }
\end{align}

\subsection{Gauge transformations}An MPS has a gauge degree of freedom, which we will exploit to turn $ | \psi(\bm \vartheta, \bm \varphi)\rangle$ into a normalized form. Let $\mc A(a_i, b_i) \to \mc A'(a_i,b_i,c_i,a_{i+1},c_{i+1}) = B(a_i,c_i) A(a_i, b_i) B^{-1}(a_{i+1}, c_{i+1})$ where 
\begin{align}
B(a_i,c_i) = \begin{pmatrix} 1 & 0  \\ 0 & \frac{c_i}{a_i} \end{pmatrix}.
\end{align}
We have also introduced variables $c_i$s that depend on $\left(\bm{a}, \bm{b} \right)$, which we will choose below. Then
\begin{align}
\mc A'(a_i,b_i,c_i,a_{i+1},c_{i+1}) = c_i
 \begin{pmatrix} 
\frac{a_i}{c_i} |0\rangle_i & \frac{b_i a_{i+1}}{c_i c_{i+1}}|1\rangle_i  \\  |0\rangle_i & 0 \end{pmatrix}.
\end{align}
Dropping the prefactor $c_i$ does not affect the nature of the state as it is just a normalization factor. Thus, let $\mathcal{A}'(a_i,b_i,c_i,a_{i+1},c_{i+1}) \to A(a_i,b_i,c_i,a_{i+1},c_{i+1}) = \mathcal{A}'(a_i,b_i,c_i,a_{i+1},c_{i+1})/c_i$. At this stage, let us choose $c_i(\bm{a},\bm{b})$ so that the condition
\begin{align}
\frac{|a_i|^2}{|c_i|^2} + \frac{|b_i|^2 |a_{i+1}|^2}{|c_i|^2 |c_{i+1}|^2} = 1
\end{align}
is satisifed.
We note that there is a solution, as we can rewrite the above condition as
\begin{align}
G_i = 1 +  \frac{F_i}{G_{i+1}} 
\end{align} 
where $F_i = {|b_i|^2 }/{|a_i|^2}$ and $G_i = |c_i|^2/|a_i|^2$. This gives a recurrence relation; writing it out we have explicitly a generalized continued fraction
\begin{align}
G_i = 1 + \cfrac{F_i}
{
1 + \cfrac{F_{i+1}}
{
1 + \cfrac{F_{i+2}}
{
\cdots
}
}
}
\end{align}
Assuming that $G_{L+1} = G_1$, the continued fraction becomes periodic, and one can write down the quadratic equation that $G_i$ obeys, so that $|c_i|^2 = G_i |a_i|^2$ can be explicitly solved in terms of $F_i$s which are each a function of $(a_i, b_i)$. However, the solution does not fix the phase of $c_i$. We can therefore fix it to be real, so that we can define real parameters $(\theta_i, \phi_i)$ so that 
 $\cos(\theta_i/2)=a_i/c_i$ and $-i e^{i\phi_i} \sin(\theta_i/2)= \frac{b_i a_{i+1}}{c_i c_{i+1}}$. One thus sees that $(\bm \vartheta, \bm \varphi)$ are related to $(\bm \theta, \bm \phi)$ by a non-local mapping. 
In particular, the angle $\theta_i$ at `site $i$', depends on the azimuthal angles $\bm \vartheta$ at all other sites.

With this choice of $c_i$s, we claim that the state 
\begin{align}
&  | \psi(\bm \theta, \bm \phi)\rangle  \equiv \frac{| \psi(\bm \vartheta, \bm \varphi)\rangle }{||| \psi(\bm \vartheta, \bm \varphi)\rangle ||} = \text{Tr}(A_1 A_2 \cdots A_L), \nonumber \\
& A_i(\theta_i, \phi_i) = \begin{pmatrix}
P_i |(\theta_i, \phi_i)\rangle & Q_i |(\theta_i, \phi_i)\rangle \\
 | 0 \rangle_i & 0
\end{pmatrix}
=
\left(\begin{array}{cc} \cos(\theta_i/2)\ket{0}_i & -i e^{i\phi_i}\sin(\theta_i/2)\ket{1}_i\\ \ket{0}_i & 0\end{array}\right)
\end{align}
where $|(\theta_i, \phi_i)\rangle = e^{i {\phi_i}/{2} }e^{i \phi_i S^z_i} e^{-i \theta_i S^x_i } |0\rangle_i $,  is now normalized, in the thermodynamic limit.


\subsection{Norm}
 To see this, let us calculate its norm explicitly. To this end we define the transfer matrix on a given site:
\begin{align}
\mc{T}(\bar{\theta},\bar{\phi},\theta,\phi)={A(\bar{\theta},\bar{\phi}) }^\dag \otimes A(\theta,\phi)
=
\left(\begin{array}{cccc}\cos(\bar{\theta}/2)\cos(\theta/2) & 0&0&e^{i(\phi-\bar{\phi})} \sin(\bar{\theta}/2)\sin(\theta/2) \\ \cos(\bar{\theta}/2) & 0 & 0 & 0 \\\cos(\theta/2) & 0 & 0 & 0 \\1 & 0 & 0 & 0\end{array}\right)
\end{align}
where the hermitian-conjugating operation $(^\dag)$ acts in an element-wise fashion on the matrix. 
Evaluating $\mc{T}(\theta,\phi,\theta,\phi)$ yields $T(\theta) = \mc{T}(\theta,\phi,\theta,\phi)$, where
\begin{align}
T(\theta)=\left(\begin{array}{cccc}\cos^2(\theta/2) & 0&0& \sin^2(\theta/2) \\ \cos(\theta/2) & 0 & 0 & 0 \\\cos(\theta/2) & 0 & 0 & 0 \\1 & 0 & 0 & 0\end{array}\right),
\label{eqn:AppendixT}
\end{align}
which does not depend on $\phi$.
%

Now, $T(x)$'s left and right eigenvectors are found to be
\begin{align}
(l_1(x)|&=\left(\begin{array}{cccc}1 & 0 & 0 & \sin^2(x/2)\end{array}\right)/(1+\sin^2(x/2))\\
(l_2(x)|&=\left(\begin{array}{cccc}-1 & 0 & 0 & 1\end{array}\right)/(1+\sin^2(x/2))\\
(l_3(x)|&=\left(\begin{array}{cccc}0 & 1 & 0 & -\cos(x/2)\end{array}\right)\\
(l_4(x)|&=\left(\begin{array}{cccc}0 & -1 & 1 & 0\end{array}\right)
\end{align}
and
\begin{align}
|r_1(x))&=\left(\begin{array}{c}1 \\\cos(x/2) \\\cos(x/2) \\1\end{array}\right),\quad 
|r_2(x))=\left(\begin{array}{c}-\sin^2(x/2) \\\cos(x/2) \\\cos(x/2) \\1\end{array}\right),\quad 
|r_3(x))=\left(\begin{array}{c}0 \\1 \\1 \\0\end{array}\right),\quad
|r_4(x))=\left(\begin{array}{c}0 \\0 \\1 \\0\end{array}\right)
\end{align}
with corresponding eigenvalues  given by
\begin{align}
\lambda_1(x)=1,\quad
\lambda_2(x)=-\sin^2(x/2),\quad
\lambda_3(x)=0,\quad
\lambda_4(x)=0.\quad.
\end{align}
Note that these eigenvectors are normalized such that $(l_i(x)|r_j(x))=\delta_{i,j}$. With this we can resolve the identity as
$
1=\sum_{k=1}^4|r_k(x))(l_k(x)|
$. 
We also will use the following notation $(l_{i}(x)|r_{j}(y))=\mc{M}_{i,j}(x,y)$ which gives the matrix 
\begin{align}
\mc{M}(x,y)=\left(\begin{array}{cccc}
1 & \frac{\sin^2(x/2)-\sin^2(y/2)}{1+\sin^2(x/2)} & 0 & 0 \\
0 & \frac{1+\sin^2(y/2)}{1+\sin^2(x/2)} & 0 & 0 \\
\cos(y/2)-\cos(x/2) & \cos(y/2)-\cos(x/2) & 1 & 0 
\\0 & 0 & 0 & 1\end{array}\right).
\end{align}
which we point out has matrix element $\mc{M}_{2,1}(x,y)=0$.
By definition, we have that
\begin{align}\mc{M}(x,y)\mc{M}(y,z)=\mc{M}(x,z),\end{align}
and also that, for all $k=1,2,3,4$,
\begin{align}
\mc{M}_{k,k}(x,y)\mc{M}_{k,k}(y,z)=\mc{M}_{k,k}(x,z).
\end{align}
Using these we are now equipped to calculate the norm of our variational state:
\begin{align}
\braket{\psi(\vec{\theta}, \bm \phi)}{\psi(\vec{\theta}, \bm \phi)} &=\Tr\left[T(\theta_1)T(\theta_2)\dots T(\theta_N) \right]\\
&=\sum_{k_1,\dots, k_N=1}^4\lr{\prod_{j=1}^{N}\lambda_{k_j}(\theta_j)}(l_{k_1}(\theta_1)|r_{k_2}(\theta_2))(l_{k_2}(\theta_2)|r_{k_3}(\theta_3))\dots(l_{k_N}(\theta_N)|r_{k_1}(\theta_1)) \\
&=\sum_{k=1}^2\lr{\prod_{j=1}^{N}\lambda_{k}(\theta_j)}=1+\prod_{j=1}^{N}(-\sin^2(\theta_j/2)).
\end{align}
Since the product of $\sin^2(\theta_j/2)$s in the r.h.s.~of the above equation generically vanishes in the thermodynamic limit, this shows that the state is normalized,
\begin{align}
\boxed{\braket{\psi(\vec{\theta}, \bm \phi)}{\psi(\vec{\theta}, \bm \phi)} = 1}.
\end{align}
This is important for the purposes of the TDVP calculations in order for the dynamics to be norm preserving.

\subsection{MPS for higher spins}
We can generalize the variational MPS that we derived above for $s = 1/2$, to higher spins, by simply taking the higer-spin analogs of both the operators and states: 
\begin{align}
&  | \psi(\bm \theta, \bm \phi)\rangle  \equiv  \text{Tr}(A_1 A_2 \cdots A_L), \nonumber \\
& A_i(\theta_i, \phi_i) = \begin{pmatrix}
P_i |(\theta_i, \phi_i)\rangle & Q_i |(\theta_i, \phi_i)\rangle \\
 | 0 \rangle_i & 0
\end{pmatrix}
\label{eqn:appendixMPS}
\end{align}
where $|(\theta_i, \phi_i)\rangle = e^{i {\phi_i s} }e^{i \phi_i S^z_i} e^{-i \theta_i S^x_i } |0\rangle_i $, $P_i = |0\rangle_i \langle 0|_i$ and $Q_i = \mathbb{I}_i - P_i$. Once again, this state is normalized in the thermodynamic limit. This can be seen easily from the fact that the transfer matrix is
\begin{align}
\mathcal{T}(\theta, \phi, \theta, \phi) = A(\theta,\phi)^\dag \otimes A(\theta,\phi) = 
\begin{pmatrix}
\langle (\theta,\phi)| P | (\theta,\phi)\rangle & 0 & 0 &   \langle (\theta,\phi)| Q | (\theta,\phi)\rangle \\
\langle (\theta,\phi)| 0 \rangle & 0 & 0 & 0 \\
\langle 0 | (\theta,\phi) \rangle & 0 & 0  & 0 \\
1 & 0 & 0 & 0
\end{pmatrix}
\label{eqn:AppendixTmatrix}
\end{align}
which similarly to the $s=  1/2$ case has a single dominant eigenvalue equal to $1$.

\section{II. TDVP Calculations}
The time-dependent variational principle generates dynamics on a variational manifold of states that is most `optimal',  a condition which  can be formulated in two generically equivalent ways: (i) the geometric principle, and (ii) the action principle. 

In the former geometrical principle, dynamics on the variational manifold is derived by continually projecting the full quantum evolution at any point in the manifold onto its tangent space, so that motion always remains within the manifold. In other words, assuming a parameterization of the manifold by $\bm z$ (in our case, $\bm z = (\bm \theta, \bm \phi)$ ), one minimizes the motion out of the tangent space, or equivalently the vector orthogonal to the tangent space,
\begin{align}
\min_{ \dot{z} } || \bm { \dot{z} } \partial_{\bm z}|\psi(\bm z)\rangle + iH |\psi(\bm z)\rangle ||.
\end{align}
This leads to the equations of motion
\begin{align}
\sum_k \langle \partial_{z_l} \psi(\bm z)| \partial_{z_k} \psi(\bm z)\rangle \dot{z}_k + i \langle \partial_{z_l} \psi(\bm z) | H | \psi(\bm z)\rangle,
\label{eqn:appendixEOM}
\end{align}
where $\langle \partial_{z_l} \psi(\bm z)| \partial_{z_k} \psi(\bm z)\rangle $ is the so-called Gram matrix. In this geometrical picture, the instantaneous error resulting from the TDVP  motion can naturally be quantified as
\begin{align}
\Gamma(\bm z) = || \bm { \dot{z} } \partial_{\bm z}|\psi(\bm z)\rangle + iH |\psi(\bm z)\rangle ||.
\end{align}
The error between the true unitary and the TDVP time evolution, is then upper bounded as $|| e^{-i H t} | \psi_0 \rangle - | \psi(\bm z(t) )\rangle || \leq  \int_0^t \Gamma(\bm z(t)) dt$, where $|\psi(\bm z(0) ) \rangle = |\psi_0\rangle$.  In a many-body system, however, since $\Gamma(\bm z)$ scales as $\sim \langle \psi(\bm z)| H^2 | \psi(\bm z)\rangle ^2 \sim L$, this is not a particularly useful bound. Instead, we will consider the normalized, intensive version of the error,
\begin{align}
\gamma(\bm z) = \Gamma(\bm z) / \sqrt{L},
\label{eqn:AppendixError}
\end{align}
where $L$ is the total number of sites, as was used in the main text. This has the interpretation of the instantaneous rate of leakage per site of the wavefunction out of the manifold.

In the latter action principle, one extremizes the action of the following classical Lagrangian:
\begin{align}
\mathcal{L} = i \langle \psi(\bm z) | \partial_{\bm z} \psi (\bm z) \rangle \dot{\bm z}  - \langle \psi (\bm z)| H | \psi (\bm z)\rangle,
\end{align}
where in the above, it is implicitly assumed that the dimensionality of the manifold is large enough to support a symplectic structure; this means that $\bm z$ must be at least even dimensional. 

We note here that the TDVP has the property that it generates classical dynamics in the phase space $\bm z$, via the Lagrangian $\mathcal{L}$ above, or equivalently, the corresponding Hamiltonian which is related via a Legendre transformation. Thus, the TDVP respects conservation laws. In particular, the energy of the system is conserved: that is, $\partial_t \langle \psi( \bm z) | H  | \psi(\bm z) \rangle = 0$, a fact that will be useful to us in simplying the following calculations.

\subsection{Geometric principle}
Let us derive the EOMs for all spin-$s$ representations using the geometric principle of the TDVP on the states $|\psi(\bm \theta, \bm \phi)\rangle$ (\ref{eqn:appendixMPS}), i.e.~by evaluating Eqn.~(\ref{eqn:appendixEOM}). As we are interested in describing the dynamics of the states $| \bm 0\rangle$ and $|\mathbb{Z}_2 \rangle$, we will further focus on the states having a two-site unit cell translational invariance, i.e.~$(\theta_{2i}, \phi_{2i}) =(\theta_e, \phi_e)$ and $(\theta_{2i+1},\phi_{2i+1})=(\theta_o,\phi_o)$. We first establish some notations:
\begin{align}
| (\theta,\phi)\rangle & \equiv P | (\theta, \phi) \rangle + Q | (\theta, \phi)\rangle  \nonumber  \\
& = x |0\rangle + Q |(\theta,\phi)\rangle,
\end{align}
which defines $x \equiv \langle 0 |  (\theta,\phi) \rangle$.  Then we have
\begin{align}
\langle (\theta, \phi) | P | (\theta, \phi) \rangle = |x|^2, \qquad \langle (\theta, \phi) | Q | (\theta, \phi) \rangle = 1 - |x|^2.
\end{align}
The one-site transfer matrix (\ref{eqn:AppendixTmatrix}) is then
\begin{align}
\mathcal{T}(\theta, \phi ) = A(\theta,\phi)^\dag \otimes A(\theta,\phi) = 
\begin{pmatrix}
|x|^2  & 0 & 0 &   1 - |x|^2 \\
x^*   & 0 & 0 & 0 \\
x & 0 & 0  & 0 \\
1 & 0 & 0 & 0
\end{pmatrix} \equiv T(x),
\end{align}
which looks similar to (\ref{eqn:AppendixT}) for $s = 1/2$.
The two site-transfer matrix, which we will use extensively in the calculations below, is given by $T(x_o, x_e) \equiv T(x_o) T(x_e)$. Its left and right eigenvectors are found to be
\begin{align}
((l_1 |&= \frac{1}{|x_o|^2 + |x_e|^2 - |x_o|^2|x_e|^2} \left(\begin{array}{cccc} |x_e|^2 & 0 & 0 & |x_o|^2(1 - |x_e|^2\end{array}\right) \\
((l_2 |&=\left(\begin{array}{cccc} 0 & 0 & 1 & -x_o\end{array}\right)\\
((l_3 |&=\left(\begin{array}{cccc}0 & 1 & 0 & -x_o^* \end{array}\right)\\
((l_4 |&= \frac{1}{|x_o|^2 + |x_e|^2 - |x_o|^2|x_e|^2}\left(\begin{array}{cccc} - |x_e|^2 & 0  & 0 & |x_e|^2\end{array}\right)
\end{align}
and
\begin{align}
|r_1 ))&=\left(\begin{array}{c}1 \\ x_o^* \\ x_o \\1\end{array}\right),\quad 
|r_2 ))=\left(\begin{array}{c}      0\\0\\1\\0                        \end{array}\right),\quad 
|r_3 ))=\left(\begin{array}{c}0\\1\\0\\0\end{array}\right),\quad
|r_4 ))=\left(\begin{array}{c}   \frac{|x_o|^2(-1+|x_e|^2)}{|x_e|^2} \\x_o^* \\ x_o \\ 1            \end{array}\right)
\end{align}
with corresponding eigenvalues  given by
\begin{align}
\lambda_1=1,\quad
\lambda_2 = 0,\quad
\lambda_3 = 0,\quad
\lambda_4 = (-1+|x_o|^2)(-1+|x_e|^2).\quad.
\end{align}

\subsubsection{Showing that $(\dot \phi_o, \dot \phi_e) = (0,0)$ and $(\phi_o, \phi_e) = (0,0)$ }

Let us now show that for the states of interest, $|\bm 0\rangle$ and $|\mathbb{Z}_2\rangle$, $(\dot \phi_o, \dot \phi_e) = (0,0)$  and therefore we have that $(\phi_o, \phi_e) = (0,0)$. This follows from evaluating the energy expectation value $\langle \psi(\bm \theta, \bm \phi) | H | \psi(\bm \theta \bm \phi)\rangle$ where $H = \Omega \sum_i \mathcal{P} S^x_i \mathcal{P}$.  To wit:
\begin{align}
h(\theta, \phi ) & \equiv A(\theta,\phi)^\dag \otimes (S^x A(\theta,\phi)) \nonumber \\
& = 
\begin{pmatrix}
\langle (\theta, \phi) | P & \langle (\theta,\phi) | Q \\
\langle 0 | & 0 
\end{pmatrix} \otimes
\begin{pmatrix}
S^x P | (\theta, \phi) \rangle &  S^x Q | (\theta,\phi) \rangle \\
S^x |  0 \rangle  & 0 
\end{pmatrix}  \nonumber \\
& = \begin{pmatrix}
0 & x^* y & x y^* & s_x - x^* y - x y^* \\
0 & 0 & y^*   & 0 \\
0 & y & 0 & 0\\
0 & 0 & 0 & 0 
\end{pmatrix}
\end{align}
where $y \equiv \langle 0 | S^x | (\theta, \phi)\rangle$ and $s_x \equiv \langle (\theta,\phi) | S^x | (\theta,\phi)\rangle$. Then,
\begin{align}
\langle \psi(\bm \theta, \bm \phi) | H | \psi(\bm \theta \bm \phi)\rangle & =
\frac{L}{2} \Omega \sum_{i=1}^4 ((l_i | h(\theta_o, \phi_o) T(x_e) + T(x_o) h(\theta_e, \phi_e) | r_i)) \nonumber \\
& = \frac{L}{2} \Omega \frac{ 
s_{x_e} |x_o|^2 + s_{x_o} |x_e|^2 + |x_o|^2 x_o^* x_e y_2 + |x_e|^2 x_o^* x_e^* y_1 + x_o x_e |x_e|^2 y_1^* + |x_o|^2 x_o x_e y_2^*
}{|x_e|^2 + |x_o|^2 - |x_o|^2 | x_e|^2} \nonumber \\
& = 
\frac{L}{2} \Omega
\frac{
2s  \cos^{4s-1}\left( \frac{\theta_o}{2}\right) \cos^{6s}\left( \frac{\theta_e}{2}\right) \sin\left( \frac{\theta_o}{2}\right) \sin(\phi_o) + (e \leftrightarrow o) 
}{\cos^{4s}\left( \frac{\theta_o}{2}\right)    + \cos^{4s}\left( \frac{\theta_e}{2}\right)  - \cos^{4s}\left( \frac{\theta_o}{2}\right) \cos^{4s}\left( \frac{\theta_e}{2}\right)   },
\end{align}
where we have used that $x = \cos^{2s}(\theta/2), y = i s \cos^{2s-1}(\theta/2) \sin(\theta/2) e^{i \phi}, s_x = s \sin(\theta) \sin(\phi)$.

Clearly, $\langle \psi(\bm \theta, \bm \phi) | H | \psi(\bm \theta \bm \phi ) \rangle= 0$ for $ (\phi_o, \phi_e) = (0,0) $. This energy expectation value equals the energies of the states $|\bm 0\rangle$ and $|\mathbb{Z}_2\rangle$, and the angles $ (\phi_o, \phi_e) = (0,0) $ encompass the states. Since the EOMs from the TDVP preserve energy expectation values, we have therefore that $(\dot \phi_o, \dot \phi_e) = (0,0)$, and we can henceforth drop all dependence on $\phi$ in our calculations, so that $|(\theta,\phi) \rangle \to | \theta \rangle = e^{-i \theta S^x } |0\rangle$.

\subsubsection{Gram matrix}
With $|(\theta,\phi) \rangle \to | \theta \rangle = e^{-i \theta S^x } |0\rangle$, let us calculate the two-by-two Gram matrix
\begin{align}
G_{\mu \nu} \equiv  \langle  \partial_{\theta_\mu} \psi(  \theta_o,   \theta_e) | \partial_{\theta_\nu} \psi(\theta_o, \theta_e) \rangle
\end{align}
where $\mu, \nu = o,e$. We have also that $G_{oe} = G_{eo} |_{\theta_o \to \theta_e, \theta_e \to \theta_o}$ and $G_{ee} = G_{oo} |_{\theta_o \to \theta_e, \theta_e \to \theta_o}$, so it suffices to calculate $G_{oe}$ and $G_{oo}$. The following objects will also be useful to us:
\begin{align}
\bar{\partial}T(x) & \equiv \partial_\theta A(\theta)^\dag \otimes A(\theta) 
=
\begin{pmatrix}
x (x^*)' & 0 & 0 & - x(x^*)' \\
(x^*)' & 0 & 0 & 0 \\
0 & 0 & 0 & 0\\
0 & 0 & 0 & 0
\end{pmatrix}
, \nonumber \\
\partial T(x) & \equiv   A(\theta)^\dag \otimes \partial_\theta A(\theta)  
=
\begin{pmatrix}
x^* x' & 0 & 0 & - x^* x' \\
0 & 0 & 0 & 0 \\
x' & 0 & 0 & 0\\
0 & 0 & 0 & 0
\end{pmatrix}
\end{align}
where 
\begin{align}
A(\theta) = \begin{pmatrix}
P |\theta\rangle & Q |\theta\rangle  \\
| 0\rangle & 0
\end{pmatrix}
\end{align}
and $(')$ refers to the derivative with respect to $\theta$.

To compute $G_{oe}$, we consider the two different cases depending on where the two derivatives $(\partial_{\theta_o}, \partial_{\theta_e})$ act: (i) within the same (two-site) unit cell or (ii) in different unit cells. For (i), we have  
\begin{align}
\sum_{i=1}^4 \lambda_i^{L/2-1} (( l_i | \bar \partial T(x_o)  \partial T(x_e) | r_i ))  = 0
\end{align}
in the thermodynamic limit, while for (ii), we also have 
\begin{align}
\sum_{i=1}^4 \sum_k (( l_i |  \left[ \bar \partial T(x_o) T(x_e)  \right] T(x_o, x_e)^k  \left[ T(x_o) \partial T(x_e) \right] T(x_o, x_e)^{L/2-k-2} | r_i )) = 0
\end{align}
in the thermodynamic limit. Thus, 
\begin{align}
\boxed{G_{oe} = G_{eo} = 0}
\end{align}

To compute $G_{oo}$, we consider also the two cases where the two derivatives act. For the case where they act on differing unit cells, we once again have
\begin{align}
\sum_{i=1}^4 \sum_k (( l_i |  \left[ \bar \partial T(x_o) T(x_e)  \right] T(x_o, x_e)^k  \left[ \partial T(x_o)  T(x_e) \right] T(x_o, x_e)^{L/2-k-2} | r_i )) = 0
\end{align}
in the thermodynamic limit. For the case where they act on the same unit cell, we have
\begin{align}
& \sum_{i=1}^4 \lambda_i^{L/2-1} (( l_i | 
\begin{pmatrix}
\langle \partial_{\theta_o} \theta_o| P | \partial_{\theta_o} \theta_o \rangle & 0 & 0 & \langle \partial_{\theta_o} \theta_o| Q | \partial_{\theta_o} \theta_o \rangle \\
0 & 0 & 0 & \\
0 & 0 & 0 & \\
0 & 0 & 0 & \\
\end{pmatrix}
 T(x_e) | r_i ))  
\nonumber \\ & = \frac{\langle \theta_o| (S^x)^2 | \theta_o \rangle }{|x_o|^2 + |x_e|^2 - |x_o|^2 | x_e|^2 } 
 =  \frac{s/2}{|x_o|^2 + |x_e|^2 - |x_o|^2 | x_e|^2 }.
\end{align}
Thus, 
\begin{align}
\boxed{G_{oo} = G_{ee} =  \frac{L}{2} \frac{s/2}{|x_o|^2 + |x_e|^2 - |x_o|^2 | x_e|^2 }}.
\end{align}

\subsubsection{Dynamical term}
We now compute the dynamical term $i \langle \partial_{\theta_\mu} \psi(\theta_o,\theta_e)|H | \psi(\theta_o, \theta_e)\rangle$. Since the state is assumed to have two-site translational invariance, $S^x$ in the Hamiltonian $H$ could act on either the odd (o) or even (e) sites of the unit cell. 

We consider the scenario where $S^x$ acts on an odd site, and where the derivative (on $\theta_o$) acts in a different unit cell. Then
\begin{align}
& \sum_{i=1}^4 \sum_k  (( l_i |   \left[ \bar \partial T(x_o) T(x_e)\right] T(x_o, x_e)^k  \left[ h(x_o, y_o) T(x_e) \right]T(x_o, x_e)^{L/2-k-2} | l_i )) \nonumber \\
& =   \frac{x_o |x_e|^2 ( - 1 + |x_e|^2) (x_o^*)'( y_o x_o^* x_e^* + x_o x_e y_o^*) }{(|x_o|^2 + |x_e|^2 - |x_o|^2 |x_e|^2 )^2}.
\label{eqn:AppendixH1}
\end{align}
Next we consider the scenario where $S^x$ acts on an odd site, while the derivative acts  within the same unit cell. We then have
\begin{align}
& \sum_{i=1}^4 \lambda_i^{L/2-1} (( l_i |
\begin{pmatrix}
0 & (x_o^*)' y_o & x_o y_o' & tt_o - x_o (y_o^*)' - x_o^* y_o \\ 
0 & 0 & (y_o^*)' & 0 \\ 
0 & 0 & 0 & 0 \\
0 & 0 & 0 & 0 
\end{pmatrix}
 T(x_e) | r_i )) \nonumber \\
& = \frac{ |x_2|^2(tt_o + y_o ( -1 + x_e^*) (x_o^*)' + x_o(-1 + x_e) (y_o^*)')  }{|x_o|^2 + |x_e|^2 - |x_o|^2 | x_e|^2 }
\label{eqn:AppendixH2}
\end{align}
where $tt_o \equiv \langle \partial_{\theta_o} \psi(\theta_o,\theta_e)| S^x | \psi(\theta_o,\theta_e)\rangle$.

Moving forward, we consider the scenario where $S^x$ acts on an even site. Similarly, there are two cases: for the case where the derivative acts in a different unit cell, we have 
\begin{align}
& \sum_{i=1}^4 \sum_k  (( l_i |   \left[ \bar \partial T(x_o) T(x_e)\right] T(x_o, x_e)^k  \left[ T(x_o)  h(x_e, y_e) \right]T(x_o, x_e)^{L/2-k-2} | l_i )) \nonumber \\
& =   -\frac{|x_o|^2 y_e x_e^* (x_o^*)'}{|x_o|^2 + |x_e|^2 - |x_o|^2 | x_e|^2 } 
+
\frac{x_o x_e(-1+|x_o|^2) x_e^*(-1 + |x_e|^2) (x_o^*)'(y_e x_o^* x_e^* + x_o x_e y_e^*) }{(|x_o|^2 + |x_e|^2 - |x_o|^2 | x_e|^2 )^2}.
\label{eqn:AppendixH3}
\end{align}
Lastly, the on-site term is 
\begin{align}
\sum_{i=1}^4 \lambda_i^{L/2-1}  ((l_i| \partial T(\theta_o) h(x_e,y_e) | r_i)) = \frac{x_o (x_o^*)' |x_e|^2 (y_e x_o^* x_e^* + x_o x_e y_e^*) }{|x_o|^2 + |x_e|^2 - |x_o|^2 | x_e|^2}.
\label{eqn:AppendixH4}
\end{align}
Thus,
\begin{align}
\boxed{i \langle \partial_{\theta_o} \psi(\theta_o,\theta_e)|H | \psi(\theta_o, \theta_e)\rangle = i \frac{L}{2} \Omega \left( (\ref{eqn:AppendixH1}) + (\ref{eqn:AppendixH2}) + (\ref{eqn:AppendixH3}) + (\ref{eqn:AppendixH4})  \right)},
\end{align}
and $i \langle \partial_{\theta_o} \psi(\theta_o,\theta_e)|H | \psi(\theta_o, \theta_e)\rangle$ is given by the above expression but with $\theta_o, \theta_e$ swapped.

\subsubsection{Equations of motion}
We now evaluate all the expressions to obtain the equations of motion
\begin{align}
& \dot{\theta}_o = G_{oo}^{-1} i \langle \partial_{\theta_o} \psi(\theta_o,\theta_e)|H | \psi(\theta_o, \theta_e)\rangle \nonumber \\
&  \dot{\theta}_e = G_{ee}^{-1} i \langle \partial_{\theta_e} \psi(\theta_o,\theta_e)|H | \psi(\theta_o, \theta_e)\rangle.
\end{align}
We use that
\begin{align}
x & = \cos^{2s}\left(\frac{\theta}{2} \right) \nonumber \\
x' & = -s \sin\left(\frac{\theta}{2} \right) \cos^{2s-1}\left(\frac{\theta}{2} \right) \nonumber \\
y &= i x' \nonumber \\
y' & = -i \frac{s}{2} (1 - s + s \cos(\theta)) \cos^{2s-2}\left( \frac{\theta}{2} \right) \nonumber \\
tt & = -i s/2
\end{align}
to obtain that
\begin{align}
\boxed{\dot{\theta}_o = \Omega\left[ 1 - \cos^{4s-2}\left(\frac{\theta_o}{2}\right)
+ 
 \cos^{4s-2}\left(\frac{\theta_o}{2}\right)  \cos^{2s}\left(\frac{\theta_e}{2}\right)
+
2 s  \cos^{6s-1}\left(\frac{\theta_o}{2}\right) \sin\left(\frac{\theta_o}{2} \right) \tan\left( \frac{\theta_e}{2} \right) \right]
 },
\nonumber\\
\boxed{\dot{\theta}_e = \Omega \left[  1 - \cos^{4s-2}\left(\frac{\theta_e}{2}\right)
+ 
 \cos^{4s-2}\left(\frac{\theta_e}{2}\right)  \cos^{2s}\left(\frac{\theta_o}{2}\right)
+
2s  \cos^{6s-1}\left(\frac{\theta_e}{2}\right) \sin\left(\frac{\theta_e}{2} \right) \tan\left( \frac{\theta_o}{2} \right) \right]
 }.
\end{align}

\subsubsection{Error calculation}
We present here the calculation of the error $\Gamma$ (which is related to $\gamma$ via (\ref{eqn:AppendixError})). We have
\begin{align}
\Gamma^2 & = \langle \psi(\theta_o, \theta_e)| H^2 | \psi(\theta_o,\theta_e) \rangle - i \sum_{\mu=o,e} \dot{\theta}_\mu \langle \psi(\theta_o, \theta_e)| H | \partial_{\theta_\mu}  \psi(\theta_o,\theta_e)\rangle + i \sum_{\mu=o,e} \dot{\theta}_\mu  \langle  \partial_{\theta_\mu} \psi(\theta_o, \theta_e)| H  | \psi(\theta_o,\theta_e)\rangle \nonumber \\
& + \sum_{\mu,\nu} \dot{\theta}_\mu \dot{\theta}_\nu  \langle  \partial_{\theta_\mu} \psi(  \theta_o,   \theta_e) | \partial_{\theta_\nu} \psi(\theta_o, \theta_e) \rangle.
\end{align}
When evaluated along the EOMs derived from the TDVP, the middle two terms vanish. Since the last term is nothing but the Gram matrix, we simply have to evaluate the first term, $\langle \psi(\theta_o, \theta_e)|H^2 | \psi(\theta_o,\theta_e)\rangle$. We only present here the final result. It is given by
\begin{align}
\langle \psi(\theta_o, \theta_e)|H^2 | \psi(\theta_o,\theta_e)\rangle & = 
\frac{L}{2}  \frac{|x_o|^2  \left( \frac{s}{2} (|x_o|^2 |x_e|^2 - 2 x_o| x_e|^2 +1 + |x_e|^2  ) + 2 \langle \theta_e| (S^x)^2 | 0 \rangle (x_o x_e - x_e) \right) }{ |x_o|^2 + |x_e|^2 - |x_o|^2 | x_e|^2  } + (o \leftrightarrow e) \nonumber \\
& + \frac{L}{2} \frac{4 |x_o|^2 |x_e|^2 \langle \theta_e|(S^x)^2 |0\rangle \langle 0 | (S^x)^2 | \theta_o \rangle }{ |x_o|^2 + |x_e|^2 - |x_o|^2 | x_e|^2  }.
\end{align}

\subsubsection{Error along the trajectory of the $|0\rangle$ state}

The error around an orbit $\mathcal{C}$ of the equations of motion generated by TDVP is defined by 
\begin{align}
\epsilon = \oint_{\mathcal{C}} \gamma dt.
\end{align}
Note that the trajectory from the state $|\bm 0\rangle$, also lies on an orbit for $s=1$ and $2$. The error along such an orbit is $\epsilon_\mathcal{C} = 1.17,1.15$ for $s = 1,2$ respectively, larger than that quoted for the orbit that $|\mathbb{Z}_2\rangle$ lives on.


\subsection{Action Principle}
\subsubsection{ Full Lagrangian}

Let us calculate the full Lagrangian
\begin{align}
\mathcal{L}(\vec{\theta},\vec{\phi}) = i \langle \psi(\vec{\theta},\vec{\phi}) | \partial_{\bm \theta} \psi (\vec{\theta},\vec{\phi}) \rangle \dot{\bm \theta} + i \langle \psi(\vec{\theta},\vec{\phi}) | \partial_{\bm \phi} \psi (\vec{\theta},\vec{\phi}) \rangle \dot{\bm \phi}  - \langle \psi (\vec{\theta},\vec{\phi})| H | \psi (\vec{\theta},\vec{\phi})\rangle,
\end{align}
on which extremizing its action will yield the TDVP equations. Let us henceforth focus on the $s=1/2$ case only, so that in what follows, $T, \mathcal T$ correspond to the appropriate $s = 1/2$ matrices defined earlier. 

We define a few useful objects: $T_\sigma(\theta)=\partial_\sigma \mc{T}|_{\bar{\theta}=\theta,\phi=\bar{\phi}}$ (for $\sigma=\bar{\theta},\bar{\phi},\theta,\phi$), which once again do not depend on $\phi$:
\begin{align}
T_{\bar\theta}=\frac{1}{2}\left(\begin{array}{cccc} -\sin(\theta/2)\cos(\theta/2) & 0&0& \cos(\theta/2)\sin(\theta/2) \\ -\sin(\theta/2) & 0 & 0 & 0 \\0 & 0 & 0 & 0 \\0 & 0 & 0 & 0\end{array}\right)
\end{align}
\begin{align}
T_\theta=\frac{1}{2}\left(\begin{array}{cccc} -\sin(\theta/2)\cos(\theta/2) & 0&0& \cos(\theta/2)\sin(\theta/2) \\ 0 & 0 & 0 & 0 \\-\sin(\theta/2) & 0 & 0 & 0 \\0 & 0 & 0 & 0\end{array}\right)
\end{align}
\begin{align}
T_{\bar\phi}=-T_{\phi}=\left(\begin{array}{cccc}0 & 0&0& -i\sin^2(\theta/2) \\ 0 & 0 & 0 & 0 \\0 & 0 & 0 & 0 \\0 & 0 & 0 & 0\end{array}\right).
\end{align}

Now, we have:
\begin{align}
\braket{\psi(\vec{\theta},\vec{\phi})}{{\partial}_{\theta_i}\psi(\vec{\theta},\vec{\phi})}&=\tr\{T(\theta_1)\dots T_{\theta_i}(\theta_i)\dots T(\theta_N)\}\\
&=\sum_{k_1,\dots, k_N=1}^4\lr{\prod_{j\neq i}\lambda_{k_j}(\theta_j)}\mc{M}_{k_1,k_{2}}(\theta_1,\theta_{2})
\dots
\mc{M}^\theta_{k_i,k_{i+1}}(\theta_i,\theta_{i+1})
\dots
\mc{M}_{k_{N},k_{1}}(\theta_{N},\theta_{1}) \\
&=\frac{1}{2}\cos(\theta_i/2)\prod_{j}(-\sin^2(\theta_j/2)).
\end{align}
Here we used again the structure of the matrix $(l_{i}(x)|T_\theta(x)|r_{j}(y))=\mc{M}_{i,j}^{\theta}(x,y)$ given by
\begin{align}
\mc{M}^\theta(x,y)=
\frac{1}{2}\left(
\begin{array}{cccc}
 0 & \frac{\sin(x/2)\cos(x/2)(\sin^2(y/2)+1)}{\sin^2(x/2)+1} & 0 & 0
   \\
 0 & -\frac{\sin(x/2)\cos(x/2)(\sin^2(y/2)+1)}{\sin^2(x/2)+1}  & 0 & 0
   \\
 0 & 0 & 0 & 0 \\
 -\sin (x/2) & \sin (x/2) \sin ^2(y/2) & 0 & 0 \\
\end{array}
\right).
\end{align}
Similarly we find 
\begin{align}
& \braket{\psi (\vec{\theta},\vec{\phi})}{{\partial}_{\phi_i}\psi (\vec{\theta},\vec{\phi})}
=\frac{i\sin^2(\theta_i/2)}{\sin ^2(\theta_i/2)+1}F_i(\vec{\theta})\\
&F_i(\vec{\theta})=\lr{1-\sum_{j\neq i}\frac{\sin^2(\theta_{j}/2)-\sin^2(\theta_{j+1}/2)}{\sin^2(\theta_{j}/2)+1}\frac{\sin^2(\theta_i/2)+1}{\sin^2(\theta_{j+1}/2)+1}\prod_{n=j+1}^{i-1}(-\sin^2(\theta_n/2)) -\prod_{j\neq i}(-\sin^2(\theta_j/2))}
\end{align}
and
\begin{align}
\braket{{\partial}_{\phi_i}\psi (\vec{\theta},\vec{\phi})}{\psi (\vec{\theta},\vec{\phi})}&=-\braket{\psi (\vec{\theta},\vec{\phi})}{{\partial}_{\phi_i}\psi (\vec{\theta},\vec{\phi})}.\end{align}

For the energy expectation value we get
\begin{align}
\bra{\psi (\vec{\theta},\vec{\phi})}\sum_i S_i^x\ket{\psi (\vec{\theta},\vec{\phi})}&
= \frac{1}{2} \sum_i \frac{\cos (\theta_{i+1}/2) \sin (\theta_i) \sin (\phi_i)}{\sin^2 (\theta_i/2)+1}F_i(\vec{\theta}).
\end{align}

In summary, in the thermodynamic limit the Lagrangian is
\begin{align}
& \boxed{ \mc{L}=\sum_{i}K_i(\vec{\theta})\lr{\sin^2(\theta_i/2)\dot\phi_i+\frac{\Omega}{2}\cos (\theta_{i+1}/2) \sin (\theta_i) \sin (\phi_i)}}, \nonumber \\
\end{align}
where
\begin{align}
K_i(\vec{\theta})&=\lr{\frac{1}{\sin^2 (\theta_i/2)+1}
+\sum_{j\neq i}\lr{\frac{1}{\sin^2(\theta_{j}/2)+1}-\frac{1}{\sin^2(\theta_{j+1}/2)+1}}\prod_{n=j+1}^{i-1}(-\sin^2(\theta_n/2))}.
\end{align}
\subsubsection{Two-site unit cell Lagrangian}
We derive the Lagrangian for a state with two-site unit cell translational invariance. This encompasses the $|\bm 0\rangle$ and $|\mathbb{Z}_2 \rangle$, in particular. 
Let $(\theta_{2i}, \phi_{2i}) =(\theta_e, \phi_e)$ and $(\theta_{2i+1},\phi_{2i+1})=(\theta_o,\phi_o)$.
\begin{align}
K_{2i}&=\frac{1}{1+\sin^2(\theta_e/2)}+
\lr{\frac{1}{1+\sin^2(\theta_e/2)}-\frac{1}{1+\sin^2(\theta_o/2)}}(1+\sin^2(\theta_o/2))\sum_{k=0}^{N/2\rightarrow\infty}\sin^{2k}(\theta_e/2)\sin^{2k}(\theta_o/2)\\
&=\frac{\cos^2(\theta_o/2)}{1-\sin^{2}(\theta_e/2)\sin^{2}(\theta_o/2)}.
\end{align}
Analogously
\begin{align}
K_{2i+1}
&=\frac{\cos^2(\theta_e/2)}{1-\sin^{2}(\theta_o/2)\sin^{2}(\theta_e/2)}.
\end{align}
Thus
\begin{align}
\boxed{ \mc{L}=\frac{\cos^2(\theta_o/2)}{1-\sin^{2}(\theta_e/2)\sin^{2}(\theta_o/2)} \left(\sin^2(\theta_e/2)\dot\phi_e+ \frac{\Omega}{2} \cos (\theta_o/2) \sin (\theta_e) \sin (\phi_e)\right)
+(e\leftrightarrow o) }.
\end{align}

%
%

\section{III. Measure and resolution of the identity }

In this section we write down the measure $\mu(\bm{\theta}, \bm{\phi})$ required for a resolution of the identity on the constrained space, which then allows for a path integral description of the system. Let us only focus on the case $s = 1/2$. The `outer product transfer matrix' is given by
\begin{align}
A(\theta,\phi) \otimes A(\theta,\phi)^\dagger =
\begin{pmatrix}
\cos^2(\theta/2) |0\rangle \langle 0 |  & i e^{-i \phi} \cos(\theta/2)\sin(\theta/2) |0\rangle\langle 1| &  i e^{i \phi} \cos(\theta/2)\sin(\theta/2) |1\rangle\langle 0|  & \sin^2(\theta/2) |1 \rangle \langle 1| \\
\cos(\theta/2) |0\rangle \langle 0| & 0 & -i e^{i \phi} \sin(\theta/2) |1\rangle \langle 0|  & 0 \\
\cos(\theta/2) |0\rangle \langle 0| & i e^{i \phi} \sin(\theta/2) |0\rangle \langle 1| & 0 &  0 \\
|0\rangle \langle 0| & 0 & 0 & 0 
\end{pmatrix}.
\end{align}
Let us postulate an ansatz for the measure to be
\begin{align}
\mu(\theta, \phi) = \frac{1}{2\pi} ( \alpha + \beta \cos(\theta)),
\end{align}
where $\phi, \theta$ are both to be integrated from $0$ to $2\pi$. Then, we have
\begin{align}
\int_0^{2\pi} \int_0^{2\pi} d \theta    d\phi \mu(\theta,\phi) A(\theta,\phi) \otimes A(\theta,\phi)^\dagger = 
\begin{pmatrix}
\frac{\pi}{2} (2 \alpha + \beta) P & 0 & 0 & \frac{\pi}{2} (2\alpha - \beta) Q \\
0 & 0 & 0 & 0\\
0 & 0 & 0 & 0\\
2\pi \alpha P & 0 & 0 & 0
\end{pmatrix}
\end{align}
where $P = |0\rangle \langle 0|$ and $Q = \mathbb{I} - P$. Choosing $\alpha = \frac{2+\sqrt{5}}{(3+\sqrt{5})\pi}, \beta = \frac{2}{(3+\sqrt{5})\pi}$ gives
\begin{align}
 \int_0^{2\pi} \int_0^{2\pi} d \theta  d\phi \mu(\theta,\phi) A(\theta,\phi) \otimes A(\theta,\phi)^\dagger = 
\begin{pmatrix}
 P & 0 & 0 & \frac{1}{\varphi} Q \\
0 & 0 & 0 & 0\\
0 & 0 & 0 & 0\\
\varphi P& 0 & 0 & 0
\end{pmatrix}
\equiv \mathbb{A}
\end{align}
where $\varphi = (1+ \sqrt{5})/2$ is the Golden Ratio. Thus, 
\begin{align}
\int\int \prod_i \mu(\theta_i, \phi_i) | \psi(\bm \theta, \bm \phi)\rangle \langle \psi(\bm \theta, \bm \phi) | & = \int\int \prod_i \mu(\theta_i, \phi_i)  \text{Tr}(A_1 A_2 \cdots A_L)  \left(  \text{Tr}(A_1 A_2 \cdots A_L) \right)^\dagger\nonumber \\
& = \int\int \prod_i \mu(\theta_i, \phi_i) \text{Tr}\left( \prod_i A(\theta_i, \phi_i) \otimes A(\theta_i,\phi_i)^\dagger \right) \nonumber \\
& = \text{Tr}\left( \mathbb{A}_1 \mathbb{A}_2 \cdots \mathbb{A}_L \right) \nonumber \\
& = P_1 P_2 \cdots P_L + Q_1 P_1 \cdots P_L + P_1 Q_1 P_2 \cdots P_L + \cdots + P_1 Q_1 P_2 Q_3 \cdots Q_L + \cdots \nonumber \\
& = \mc{P},
\end{align}
the identity operator on the constrained space. This is because the trace of the product of $\mathbb{A}_i$s generates an equal weight linear combination of all possible products of local projectors onto`$0$'s and `$1$'s states which are consistent with the constraints. Thus, we have the resolution of the identity and the measure
\begin{align}
\boxed{ \int\int \mu(\bm \theta,  \bm \phi) | \psi(\bm \theta, \bm \phi)\rangle \langle \psi(\bm \theta, \bm \phi) |  = \mathcal{P}, \text{ where }  \mu(\bm \theta, \bm \phi) \equiv  \prod_i \mu(\theta_i,\phi_i ) }. 
\end{align}
The path integral over the `Gutzwiller projected' states then follows, with the full Lagrangian derived earlier.

\section{IV. Thermalization in the constrained space}
We derive in this section what it means to thermalize (to infinite temperature) in the constrained Hilbert spaces that the constrained spin models are defined in.  Consider a pure state $|\psi\rangle$ with zero energy $E = \langle \psi | H | \psi \rangle = 0$. The corresponding Gibbs ensemble that gives rise to a similar energy expectation value would be the infinite temperature ensemble, i.e.
\begin{align}
\frac{1}{Z} \text{Tr}(H e^{-\beta H} ) |_{\beta = 0} = 0,
\end{align}
since the spectra of the models are all particle-hole symmetric.  In the above, the trace is over states in the constrained Hilbert space. Thus, if the system does thermalize beginning from the state $|\psi\rangle$, the eigenstate thermalization hypothesis (ETH) states that the long-term expectation value of any local observable can be evaluated within the infinite-temperature Gibbs ensemble, namely
\begin{align}
\lim_{T \to \infty} \frac{1}{T} \int_0^T dt \langle \psi | O(t) | \psi \rangle  = \frac{1}{\mathcal{D}} \text{Tr}(O),
\end{align}
where $\mathcal{D}$ is the dimension of the Hilbert space.

 Note that $\text{Tr}(O)/\mathcal{D}$ can be estimated by taking the expectation value within a random vector $|\psi_r\rangle$, i.e.~$\langle \psi_r | O | \psi_r \rangle$. 
We can therefore evaluate the thermal expectation value of an operator $O$ by explicitly constructing the expected reduced density matrix of a random vector on the support of   $O$, while accounting for the global boundary conditions, which we will demonstrate below. In what follows, we will use periodic boundary conditions, but a similar calculation for systems with open boundary conditions can be straightforwardly performed. Note that the choice of boundary conditions will lead to very different thermal values, unlike in the case of normal, unconstrained spin systems in which bulk properties are insensitive to boundary conditions.

Let us consider first the case of spin $s = 1/2$ and an operator that acts on only one site, for example $S^z_1$ on site $1$. Now, a random vector can be decomposed in the product state basis,
\begin{align}
|\psi_r\rangle = c_1 | 0 0 1 \cdots \rangle + c_2 | 0 1 0 \cdots \rangle + \cdots +  c_i | 1 0 0 \cdots \rangle + \cdots. 
\end{align}
The ratio of the probabilities that the number of times `$0$' appears at site $1$ to the the number of times `$1$' appears, is $(1 + \varphi)/1$, where $\varphi = (1+ \sqrt{5})/2$ is the Golden Ratio. 
This can be derived by counting the number of states conditioning that the first site is $0 (1)$, while enforcing that the boundary conditions are respected, assuming that the rest of the system is infinitely large.
Thus, the expected reduced density matrix is given by
\begin{align}
\rho_1 = \frac{1}{Z} \left( |0\rangle \langle 0| + \frac{1}{1 + \varphi} |1 \rangle \langle 1 \rangle  \right)
\end{align}
where $Z = (2+\varphi)/(1+\varphi)$. The infinite-temperature value of $S^z_1$ is then
\begin{align}
\frac{1}{\mathcal{D}} \text{Tr}(S^z_1) = \frac{1}{Z}\left( -\frac{1}{2} + \frac{1}{1 + \varphi} \frac{1}{2} \right) = -\frac{1}{2} \frac{\varphi}{2 + \varphi} \approx -0.2236,
\end{align}
which agrees with the numerically observed value that the $|\bm{0}\rangle$ state equilibriates to. The above calculation also tell us that the expected entanglement entropy (EE) of a random vector over one site is 
\begin{align}
S_1 = -\text{Tr}(\rho_1 \log_2 \rho_1) \approx 0.8505.
\end{align}
Note that this is {\it not} the maximal value of entanglement possible, which would be $\max(S_1) = 1$.

A slightly more non-trivial example would involve the density matrix on three sites. The expected reduced density matrix of a random vector is  
\begin{align}
\rho_3 = \frac{1}{Z} \left( |000\rangle\langle 000| + |010\rangle \langle 010| +  \frac{1}{\varphi} |00\rangle \langle 001 | + \frac{1}{\varphi} | 100\rangle \langle 100 | + \frac{1}{1+\varphi} |101\rangle\langle 101|  \right).
\end{align}

As mentioned, the generalization to the case of spin-$s$ systems is straightforward. Focusing on a single site, the ratio of the number of times `$0$' appears to, `$1$', `$2$', $\cdots$, `$2s$', is $1 + r$ where $ r = (1 + \sqrt{1 + 8s})/4s$. Thus the reduced density matrix is
\begin{align}
 \rho_1 = \frac{1}{Z} \left( |0\rangle\langle 0 | + \frac{1}{1 + r } (\mathbb{I} - |0\rangle \langle 0| ) \right) 
\end{align}
where $Z = (2+r)/(1+r)$. The expectation value of $S^z_1$ is then
\begin{align}
\boxed{ \frac{1}{\mathcal{D}} \text{Tr}(S^z_1) = \frac{1}{Z} \left( -s + \frac{1}{1+r} \frac{s}{2s} \right) = -s \frac{-1 + 4s + \sqrt{1 + 8s}}{1 + 8s + \sqrt{1 + 8s} } }.
\end{align}
This evaluates to $-0.2236, -0.5, -1.053$, for $s = 1/2, 1, 2$ respectively, which agree with the values that the $|\bm{0}\rangle$ state equilibriates to in all cases.

\section{V. TDVP calculations for a deformed Hamiltonian}
In Ref.~[1], it has been noted that, for the case of $s=1/2$, the atypical thermalization dynamics of the $\ket{\mathbb{Z}_2}$ initial state can be enhanced  by the addition of a  suitable small perturbation. More specifically, it has been numerically demonstrated that the dynamics of $\ket{\mathbb{Z}_2}$ under the Hamiltonian
\begin{align}
H = \Omega \sum_i \mathcal{P} S_i^x \mathcal{P} + h  \sum_i   \left( P_{i-1} S_i^x P_{i+1} S_{i+2}^z +  S_{i-2}^z P_{i-1} S_i^x P_{i+1}   \right),
\end{align}
exhibits  periodic oscillations for a longer duration of time for a certain small value of $h$, despite that the initial energy density still corresponds to that of the infinite temperature ensemble. Note that the term represented by $h$ respects the constraint $\mathcal{P}$.
In this section, we repeat our TDVP analysis for this Hamiltonian and show that the enhancement of the atypical dynamics can be quantified within our calculations.
For brevity, we omit the details of the derivations, which is similar to calculations presented above, and simply present the results.

\begin{figure}[t] \includegraphics[width=0.96\textwidth]{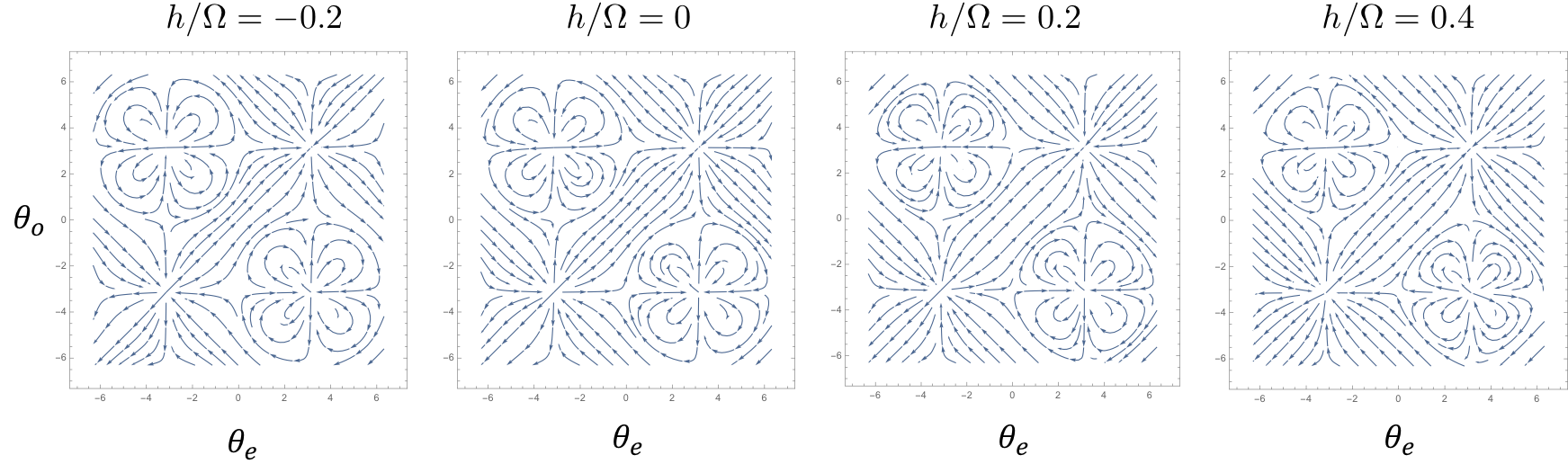}
\caption{Flow diagrams  of $\dot{\theta}_e(t), \dot{\theta}_o(t)$  for the Hamiltonian with various perturbation strengths $h/\Omega$. For a range of $h/\Omega$, the closed trajectory is sill present. 
}
\label{sfig:FlowDiagram_deformed}
\end{figure}

Similar to the previous case, we use the variational many-body wavefunction which has a  matrix product state representation:
\begin{align}
&  | \psi(\bm \theta, \bm \phi)\rangle  \equiv \frac{| \psi(\bm \vartheta, \bm \varphi)\rangle }{||| \psi(\bm \vartheta, \bm \varphi)\rangle ||} = \text{Tr}(A_1 A_2 \cdots A_L), \nonumber \\
& A_i(\theta_i, \phi_i) = \begin{pmatrix}
P_i |(\theta_i, \phi_i)\rangle & Q_i |(\theta_i, \phi_i)\rangle \\
 | 0 \rangle_i & 0
\end{pmatrix}
=
\left(\begin{array}{cc} \cos(\theta_i/2)\ket{0}_i & -i e^{i\phi_i}\sin(\theta_i/2)\ket{1}_i\\ \ket{0}_i & 0\end{array}\right).
\end{align}
It can be readily shown that $\phi_i = 0$ for our $\ket{\mathbb{Z}_2}$ initial state and that $\phi_i$ remains zero over time evolution within the variational manifold.
Therefore, we focus on the effective equations of motions for the parameters $\{ \theta_i\}$.
Using two-site translational invariance, we only need to consider the dynamics of two parameters $\theta_e$ and $\theta_o$. Based on the geometric principle, we compute the effective equations of motion:
\begin{align}
\dot \theta_e(t) =& \Omega \sec(\theta_o/2) \left(\cos^2(\theta_o/2) +\cos^2(\theta_e/2)\sin(\theta_e/2) \sin(\theta_o/2)\right)\nonumber \\
& + h \sec(\theta_o/2) \left( \cos(\theta_e) \cos^2(\theta_o/2)+ \cos^2(\theta_e/2) \cos(\theta_o) \sin(\theta_e/2) \sin(\theta_o/2) \right) \nonumber \\
\dot \theta_o(t) =& \Omega \sec(\theta_e/2) \left(\cos^2(\theta_e/2) +\cos^2(\theta_o/2)\sin(\theta_o/2) \sin(\theta_e/2)\right)\nonumber \\
& + h \sec(\theta_e/2) \left( \cos(\theta_o) \cos^2(\theta_e/2)+ \cos^2(\theta_o/2) \cos(\theta_e) \sin(\theta_o/2) \sin(\theta_e/2) \right).
\end{align}
These equations of motion still support a periodic orbit $\mathcal C$ as long as $h$ remains small as seen in Fig.~\ref{sfig:FlowDiagram_deformed}.

\begin{figure}[t] \includegraphics[width=0.8\textwidth]{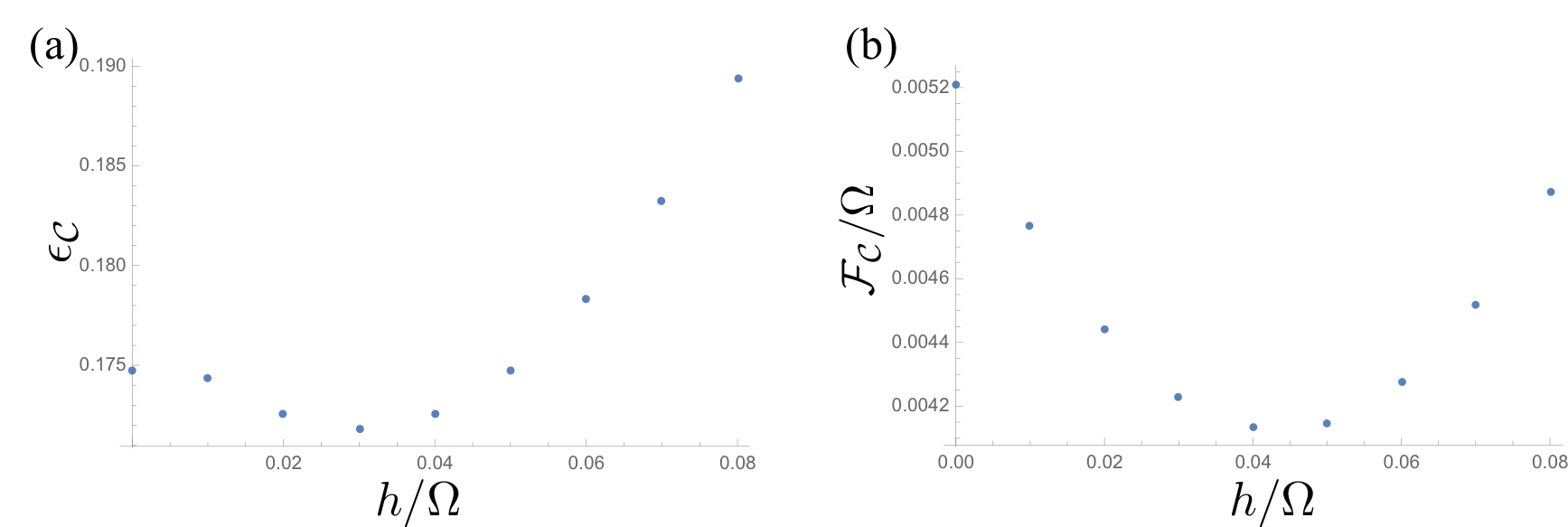}
\caption{(a) Integrated error $\epsilon_\mathcal{C}$ of the TDVP orbit $\mathcal{C}$ as a function of $h$. (b) The fluctuation $\mathcal{F}_\mathcal{C}$ of exact time evolution around the TDVP orbit $\mathcal{C}$ as a function of $h$. We find that the minimum fluctuation is achieved when $h/\Omega \approx 0.045$. 
}
\label{sfig:error_and_fluctuations}
\end{figure}

Similarly, one can also compute the integrated error $\epsilon_\mathcal{C} \equiv \oint_\mathcal{C} \gamma dt$ of the closed orbit for different values of $h$. As shown in Fig.~\ref{sfig:error_and_fluctuations}(a), we find that the error is minimized when the perturbation strength $h/\Omega$ is finite. Another important quantity that is closely related the error is the integrated ``fluctuation'' of the exact state evolution around the orbit:
\begin{align}
\mathcal{F}_\mathcal{C} \equiv \oint_\mathcal{C} \gamma^2 dt.
\end{align}
Fig.~\ref{sfig:error_and_fluctuations}(b) shows the normalized $\mathcal{F}_\mathcal{C}$ as a function of the perturbation strength $h/\Omega$. We find that the minimum fluctuation occurs at $h/\Omega \approx 0.045$. Interestingly, this value is in quantitative agreement with the optimal perturbation strength that renders the model most `integrable'-looking, with enhanced strength and duration of oscillations, as studied in Ref.~[1].
